\documentclass[aps,prd,superscriptaddress,A4paper,preprintnumbers,notitlepage,nofootinbib]{revtex4-2}
\usepackage{lmodern}
\usepackage{lineno}
\usepackage{lipsum}
\usepackage{url}
\usepackage{geometry}
\usepackage{graphicx}
\usepackage{amsmath}
\usepackage{amssymb}
\usepackage{latexsym}
\usepackage{amsfonts}
\usepackage[hidelinks]{hyperref}
\usepackage{mathtools}
\usepackage{multirow}
\usepackage{sidecap}
\usepackage{soul}
\usepackage{braket}
\usepackage{slashed}
\usepackage{booktabs}
\usepackage{bbm}
\usepackage{bm}
\def\nostrocostruttino#1\over#2{\mathrel{\mathop{\kern 0pt \rlap
{\hbox{$#1$}}} \hbox{\kern-.125em $#2$}}}

\usepackage{xcolor}

\DeclareUnicodeCharacter{2212}{-}

\begin{document}

\title{Transverse $\Lambda$ polarization in $e^+e^-$ annihilations
and in SIDIS processes  at the EIC within TMD factorization}
\author{Umberto D'Alesio}
\email{umberto.dalesio@ca.infn.it}
\affiliation{Dipartimento di Fisica, Universit\`a di Cagliari, Cittadella Universitaria, I-09042 Monserrato (CA), Italy}
\affiliation{INFN, Sezione di Cagliari, Cittadella Universitaria, I-09042 Monserrato (CA), Italy}
\author{Leonard Gamberg}
\email{lpg10@psu.edu}
\affiliation{Division of Science, Penn State Berks, Reading, PA 19610, USA }
\author{Francesco Murgia}
\email{francesco.murgia@ca.infn.it}
\affiliation{INFN, Sezione di Cagliari, Cittadella Universitaria,
I-09042 Monserrato (CA), Italy}
\author{Marco Zaccheddu}
\email{marco.zaccheddu@ca.infn.it}
\affiliation{Dipartimento di Fisica, Universit\`a di Cagliari, Cittadella Universitaria, I-09042 Monserrato (CA), Italy}
\affiliation{INFN, Sezione di Cagliari, Cittadella Universitaria, I-09042 Monserrato (CA), Italy}

\date{\today}

\begin{abstract}

We present a phenomenological study on the role of charm contribution and $SU(2)$ isospin symmetry in the extraction of the $\Lambda$ polarizing fragmentation functions  from $e^+e^- \to \Lambda^\uparrow (\bar\Lambda^\uparrow) \,h + X$ annihilation processes. We  adopt the well-established transverse-momentum-dependent factorization formalism, within the Collins-Soper-Sterman evolution scheme at next-to-leading logarithm accuracy, carefully exploiting the role
of the nonperturbative component of the polarizing fragmentation function.  We then discuss the impact of these results on the predictions for transverse $\Lambda$, $\bar{\Lambda}$  polarization in semi-inclusive deep inelastic scattering  processes at typical energies of the future Electron-Ion Collider.

\end{abstract}
\date{\today}
\maketitle

\section{Introduction}
\label{1_Intro}

The study of the fragmentation mechanism of partons into hadrons within the field theoretic framework of  quantum chromodynamics  (QCD), along with factorization theorems, which connect perturbative parton dynamics to universal hadron fragmentation functions, is
fundamental to unfolding the quark and gluon structure of hadrons.
When one includes also spin and its correlations with intrinsic transverse momentum the information one can extract is much richer and the description is more complete. This can be achieved, for instance, by studying the spontaneous transverse $\Lambda$ polarization in processes where factorization theorems, in terms of transverse momentum dependent distributions (TMDs), hold. We refer, in particular, to double-hadron production in $e^+e^-$ annihilation and semi-inclusive deep inelastic scattering (SIDIS) processes~\cite{collins_2011,Ji:2004xq,Ji:2004wu}. 
These are characterized by the presence of two ordered energy scales, a small one (the transverse momentum unbalance of the two hadrons in $e^+e^-$ processes or the transverse momentum of the final hadron in SIDIS)  and a large one, the virtuality of the exchanged photon.

We emphasize that the understanding of the  transverse $\Lambda$ polarization, originally measured in inclusive \emph{unpolarized} proton-proton and proton-nucleus collisions in the late 70's~\cite{Bunce:1976yb,Schachinger:1978qs,Heller:1978ty,Erhan:1979xm,Lundberg:1989hw,Ramberg:1994tk,Abt:2006da}, still represents a challenging problem in hadron physics.
One of the earliest attempts to describe this phenomenon within a phenomenological model was presented in Ref.~\cite{Anselmino:2000vs} and further extended to SIDIS processes in Ref.~\cite{Anselmino:2001js}.
Recently, experimental data collected by the Belle Collaboration~\cite{Belle:2018ttu}, for the transverse $\Lambda,\bar\Lambda$ polarization in almost back-to-back
two-hadron production in $e^+e^-$ processes, has triggered a renewed interest in the subject matter.  Preliminary studies within a simplified TMD model at fixed scale
were discussed in Refs.~\cite{DAlesio:2020wjq,Callos:2020qtu}.
Moreover, a series of phenomenological analyses within the TMD factorization framework adopting the Collins-Soper-Sterman (CSS) approach~\cite{Collins:1981uk,Collins:1981va, Collins:1984kg} has been carried out~\cite{Gamberg:2021iat,Kang:2021kpt,Li:2020oto,Chen:2021hdn, DAlesio:2022brl}.
The general TMD formalism, following the Lorentz decomposition or the helicity approach, was developed and presented in Refs.~\cite{Boer:1997mf,Pitonyak:2013dsu,DAlesio:2021dcx} for $e^+e^-$ processes, and in Refs.~\cite{Mulders:1995dh, Bacchetta:2006tn, Anselmino:2011ch} for SIDIS.

One of the main goals of these phenomenological studies, besides the description of data, is the extraction of the polarizing fragmentation function (pFF) for $\Lambda$ hyperons, that provides  information on the  correlations between the intrinsic transverse momentum in the parton-to-hadron fragmentation process and the final hadron polarization. In this respect, this TMD function represents a window towards a deeper understanding of the nonperturbative fragmentation mechanism when also spin-polarization effects are taken into account.

In this paper, that represents a natural extension of Ref.~\cite{DAlesio:2022brl}, we reanalyze Belle data for the transverse $\Lambda$ polarization limiting
this study to the associated production case, and  paying special attention to two issues,  mentioned in our previous work
that here we will study in depth: namely, the role of the $SU(2)$ isospin symmetry (see also Refs.~\cite{Chen:2021hdn,Chen:2021zrr}) and the charm contribution in the fragmentation of $\Lambda$ hyperons.

We consider three different scenarios, discussing their statistical significance in the data description and the difference in the  extracted polarizing fragmentation functions. We then employ these results to give predictions for the same observable in SIDIS processes at the energies and kinematics typical of the future Electron-Ion Collider (EIC). We will show how new measurements could help in disentangling among the different scenarios. The role of intrinsic charm in the proton~\cite{Brodsky:1980pb,Brodsky:2015fna} will be also addressed.

This analysis will allow us to check, at the same time, other fundamental issues, like the universality of the TMD fragmentation functions and their QCD evolution with the energy scale.

The paper is organized as follows: in Sec.~\ref{2_formalism} we present the formalism and the cross sections for the production of a transversely polarized spin-$1/2$ hadron in $e^+e^-$ collisions, in association with a light hadron, and in semi-inclusive deep inelastic scattering processes. The main results are then employed in the phenomenology part in Sec.~\ref{4_phenomenology}, where we discuss the role of the charm quark contribution and the issue of $SU(2)$ isospin symmetry in the re-analysis of Belle data \cite{Belle:2018ttu}. Estimates for the transverse $\Lambda/\bar{\Lambda}$ polarization in $e^+e^-$ collisions and in SIDIS processes, at different center of mass energies, are presented with particular focus on how these are influenced by the choice of the pFF parametrization and of the nucleon PDF set. Lastly, in Sec.~\ref{5_conclusions} we collect our concluding remarks. 
\section{Formalism}
\label{2_formalism}

In this section, we briefly recall the formalism for the production of a transversely polarized spin-1/2 hadron in $e^+e^-$ annihilation processes, in association with an unpolarized light-hadron, and in semi-inclusive deep inelastic scattering processes. The main equations will be used in the following section to study the production of transversely polarized $\Lambda$ hyperons in both processes.

\subsection{Double-hadron production in $e^+e^-$ processes}

We start considering double-hadron production in $e^+e^-$ collisions:  
\begin{equation}
 e^+(l_{e^+})\,e^-(l_{e^-})\to h_1(P_1,S_1)\, h_2(P_2) +X   \,,
\end{equation}
where $h_1$ is a spin-1/2 hadron, with momentum $P_1$, spin-polarization vector $S_{1}$ and mass $M_{1}$, while $h_2$ is a light unpolarized hadron with momentum $P_2$ (we will neglect its mass), and they are produced almost back-to-back in the center-of-mass frame of the incoming leptons. For more details we refer the reader to Refs.~\cite{DAlesio:2022brl, Boer:1997mf}.

\begin{figure}[!th]
\centering
\includegraphics[width=12cm]{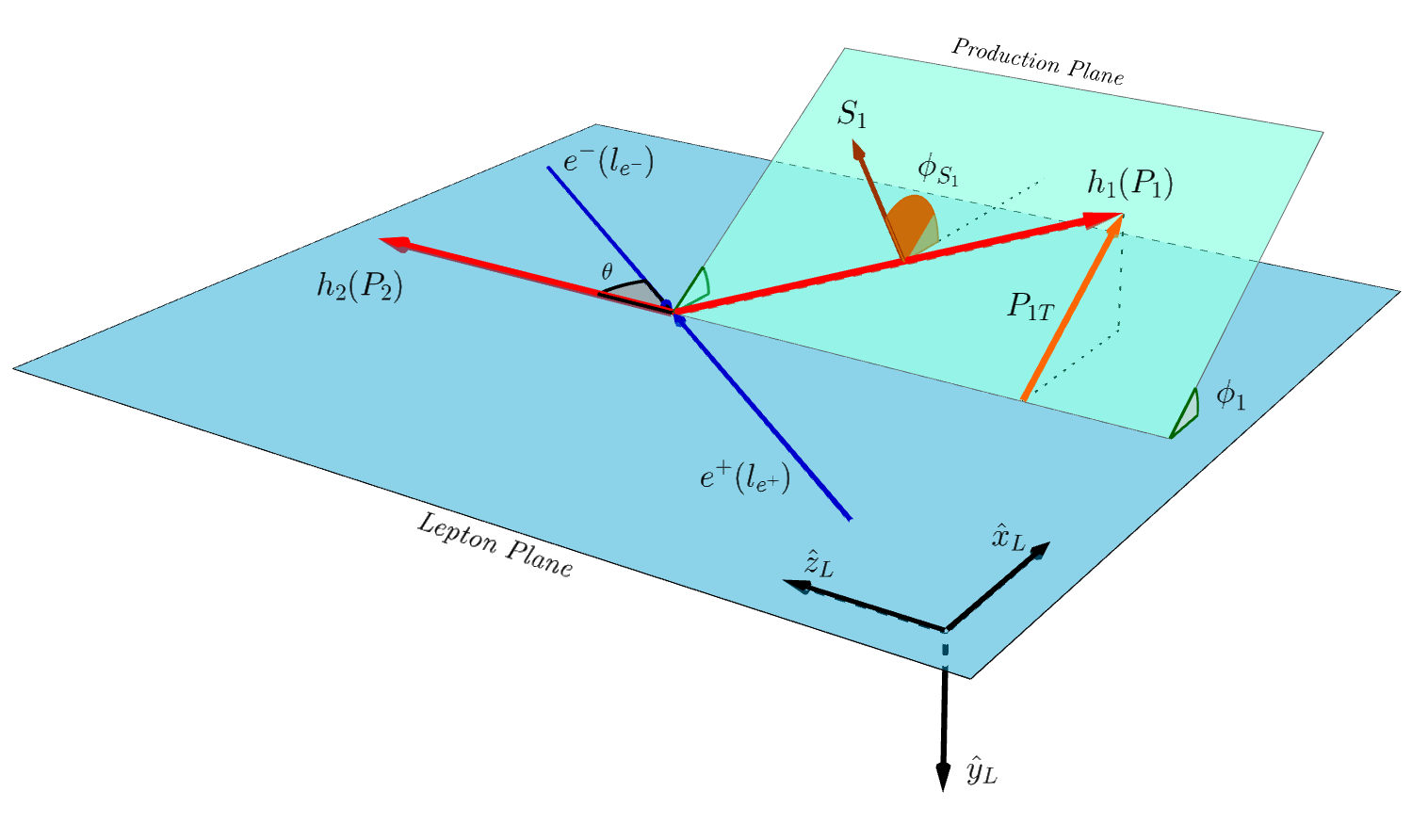}
    \caption{Kinematics for the process $e^+e^-\to h_1\, h_2 +X$ in the hadron-frame configuration.}
    \label{fig:kin_epem}
\end{figure}
In Fig.~\ref{fig:kin_epem} we show the  kinematics of the process in the hadron-frame configuration, where we fix the momentum of the second hadron, $h_2$, along the $\hat{z}_L$ axis, while the first one, $h_1$, moving in the opposite hemisphere, has a small transverse momentum $\bm{P}_{1T}$ with respect to the second hadron direction.

From the theoretical point of view, it is however more convenient to adopt a different frame, where the two hadrons are exactly back to back, along a new $\hat{\bm{z}}$ axis, and the hadron transverse unbalance ($\bm{P}_{1T}$) is now carried out by the virtual photon. In this frame, the differential cross section can be expressed, neglecting terms not relevant in the present study, as~\cite{Boer:1997mf, Callos:2020qtu, DAlesio:2021dcx} 
\begin{equation}
     \frac{d\sigma^{e^+e^-\to h_1(S_1) h_2\, X}}{2 dy\, dz_{h_1}dz_{h_2}d^2\bm{q}_{T}} = \sigma^{e^+e^-}_0\bigg[ F_{UU} - |S_{1T}|\sin(\phi_1 - \phi_{S_1}) F^{\sin(\phi_1 - \phi_{S_1})}_{TU} + \cdots \bigg]\,,   
\label{eq:xsec2h}
\end{equation}
where $\phi_{S_1}$ is the azimuthal angle of the spin of the hadron $h_1$. Here $\bm{q}_{T}$ is the transverse momentum of the virtual photon (of momentum $q$), related to the transverse momentum of the hadron $h_1$ as $\bm{P}_{1T} = - z_1 \bm{q}_T$, being $z_1$ its light-cone momentum fraction, defined for both hadrons as 
\begin{equation}
    z_1 = \frac{P^-_{1}}{p_q^-}, \qquad z_2 = \frac{P^+_{2}}{p_{\bar q}^+}\,, 
\label{eq:light_con_def}    
\end{equation}
where $p_q$ and $p_{\bar q}$ are the four-momenta of the quark and the antiquark fragmenting into the hadron $h_1$ and $h_2$, carrying a transverse momentum $\bm{k}_\perp$ and $\bm{p}_\perp$ with respect to the parent quark momenta, respectively. 

The two scaling variables in Eq.~(\ref{eq:xsec2h}), $z_{h_1}$, $z_{h_2}$, are the usual invariants (energy fractions), related to the light-cone momentum fractions as
\begin{equation}
 z_{h}  = \frac{2P_h\cdot q}{ Q^2} = \frac{2E_h}{Q} \simeq z \Bigg( 1 + \frac{M^2_{h}}{ z^2 {Q}^2}\Bigg)\,,
\label{eq:en_fract}    
\end{equation}
where $Q$ is the center-of-mass energy of the process, $Q^2 = q^2$, 
and where in the last relation we have neglected terms of the order $\mathcal{O}(\bm{k}_\perp^2/(zQ)^2)$. Another scaling variable, usually adopted in phenomenological analyses, is the hadron momentum fraction
\begin{equation}
    \begin{split}
        z_{p}  =\frac{2|\bm{P}_h|}{Q} \simeq  z \Bigg( 1 - \frac{M^2_{h}}{ z^2 {Q}^2}\Bigg)\,.
    \end{split}
\label{eq:long_fract}    
\end{equation}
Notice that since for the light hadron $h_2$ we neglect its mass, in the following we will use $z_2=~z_{h_2}=~z_{p_2}$, within this approximation.

The remaining variable is the fraction $y=P_2\cdot l_{e^+} /P_2\cdot q$, related to the polar angle $\theta$ in the hadron frame (see Fig.~\ref{fig:kin_epem}). Lastly we have 
\begin{equation}
    \sigma^{e^+e^-}_0 = \frac{3\pi\alpha^2}{Q^2}\big[y^2 + (1-y)^2\big] \,.
\label{eq:sigma_0}    
\end{equation}

In Eq.~(\ref{eq:xsec2h}), the $F$ terms are convolutions of two fragmentation functions, where the subscripts denote the polarization states of, respectively, the first and the second hadron ($U$ = unpolarized, $T$ = transversely polarized).
These have the following expressions~\cite{DAlesio:2022brl,Boer:1997mf, Callos:2020qtu}:
\begin{eqnarray}
F_{UU} & = & z_{p_1}^2z_{p_2}^2\mathcal{H}^{(e^+e^-)}(Q) \,\mathcal{F}[D_1 \bar{D}_1] \,,\\
F^{\sin(\phi_1 - \phi_{S_1})}_{TU} &=& z_{p_1}^2z_{p_2}^2\mathcal{H}^{(e^+e^-)}(Q)\, \mathcal{F}\bigg[\frac{\hat{\bm{h}}\cdot \bm{k}_T}{M_{1}}D^{\perp}_{1T} \bar{D}_1\bigg]\,, 
\label{eq:main_pol_conv_epem}    
\end{eqnarray}
where $\mathcal{H}^{(e^+e^-)}(Q)$ is the hard scattering part 
for the massless on-shell process  $e^+e^-\to q \bar{q}$ (normalized to one at leading order), at the center-of-mass energy $Q$, 
$D_1(z,{k}_{\bot})$ is the unpolarized TMD fragmentation function (FF) and $D^{\perp}_{1T}(z,{k}_{\bot})$ is the polarizing FF, with 
$\hat{\bm{h}}= \bm{P}_{1T}/|\bm{P}_{1T}|$ and $\bm{k}_T = -\bm{k}_{\bot}/z_{p_1}$ (and similarly $\bm{p}_T = -\bm{p}_{\bot}/z_{p_2}$), where $\bm{k}_T$ ($\bm{p}_T$) is the transverse momentum of the quark (antiquark) with respect to the hadron $h_1$ ($h_2$) direction of motion. The $\mathcal{F}$ are proper convolutions of TMD-FFs, defined as follows:
\begin{equation}
    \mathcal{F}[\omega  D \bar{D}] = \sum_q e^2_q \int d^2\bm{k}_T d^2\bm{p}_T \, \delta^{(2)}(\bm{k}_T + \bm{p}_T - \bm{q}_T) \,\omega(\bm{k}_T, \bm{p}_T) D(z_1,\bm{k}_{\bot}) \bar{D}(z_2,\bm{p}_{\bot}) \,,
\label{eq:conv_general}    
\end{equation}
where $\omega$ is a suitable weight factor depending on the two transverse momenta and $D$ and $\bar{D}$ are the TMD-FFs.

In order to employ the Collins-Soper-Sterman (CSS) evolution equations, it is useful to write the convolutions in the conjugate $\bm{b}_T$-space: 
\begin{equation}
F_{UU} = z_{p_1}^2z_{p_2}^2\mathcal{B}_0 \Big[\widetilde{D}_1 \widetilde{\bar{D}}_1\Big]  =
    z_{p_1}^2z_{p_2}^2\sum_q e^2_q \int \frac{d b_T}{2 \pi} \, b_T J_0(b_T\, q_T) \widetilde{D}_1(z_1,b_T) \widetilde{\bar{D}}_1(z_2,b_T)\,, 
\end{equation}
\begin{eqnarray}
 F^{\sin(\phi_1 - \phi_{S_1})}_{TU} & = & 
  M_{1} z_{p_1}^2z_{p_2}^2 \mathcal{B}_1 \Big[\widetilde{D}^{\perp  (1)}_{1T} \widetilde{\bar{D}}_1\Big] \nonumber\\
    & = & 
    M_1 z_{p_1}^2z_{p_2}^2\sum_q e^2_q \int \frac{d b_T}{2 \pi} \,b^2_T J_1(b_T\,q_T)  \widetilde{D}^{\perp  (1)}_{1T}(z_1,b_T) \widetilde{\bar{D}}_1(z_2,b_T)\,, 
\end{eqnarray}
where 
$\widetilde{D}_1(z_1,b_T)$ is the Fourier transform of the unpolarized FF,  $\widetilde{D}^{\perp  (1)}_{1T}(z_1,b_T) $ is the first moment of the polarizing fragmentation function in $\bm{b}_T$-space, and $J_i$ is the Bessel function of the first kind of $i$-th order. Notice that we have already used $\mathcal{H}^{(e^+e^-)}(Q)=1$ and all light-cone momentum fractions have to be properly understood in terms of the corresponding energy fractions, $z_h$. 

After solving the CSS evolution equations, as discussed in Refs.~\cite{collins_2011,DAlesio:2022brl}, the convolutions can be written again as:
\begin{eqnarray}
\mathcal{B}_0 \Big[\widetilde{D}_1 \widetilde{\bar{D}}_1\Big] &=&  
\frac{1}{z^2_1 z^2_2} 
\sum_q e^2_q\int \frac{d b_T}{2 \pi} \, b_T J_0(b_T\, q_T) \, d_{h_1/q}(z_1; \bar{\mu}_b) \, d_{h_2/\bar{q}}(z_2; \bar{\mu}_b)\nonumber \\
    &\times& M_{D_1}(b_c(b_T),z_1) \, M_{D_2}(b_c(b_T),z_2)
    \>e^{\!\!-g_K(b_c(b_T);b_{\text{max}})\ln{\big(\frac{Q^2 z_1 z_2}{M_{1} M_{2}}\big)}- S_{\rm pert}(b_*;\bar{\mu}_{b})}\,,
\label{B0_full_pert}
\end{eqnarray}
\begin{eqnarray}
    \mathcal{B}_1 \Big[\widetilde{D}^{\perp  (1)}_{1T} \widetilde{\bar{D}}_1\Big]
    &=& 
  \frac{1}{z^2_1 z^2_2}
    \sum_q e^2_q\int \frac{d b_T}{2 \pi} \, b^2_T J_1(b_T\, q_T) {D}^{\perp  (1)}_{1T} (z_1;\bar{\mu}_b) \, d_{h_2/\bar{q}}(z_2; \bar{\mu}_b) \nonumber\\
    &\times& M^{\perp}_{D_1}(b_c(b_T),z_1) \, M_{D_2}(b_c(b_T),z_2)\>e^{\!\!-g_K(b_c(b_T);b_{\text{max}})\ln{\big(\frac{Q^2 z_1 z_2}{M_{1} M_{2}}\big)}- S_{\rm pert}(b_*;\bar{\mu}_{b})} \,,
\label{B1_full_pert}
\end{eqnarray}
where the $d_{h/j}$'s are the $p_\perp$-integrated unpolarized fragmentation functions. $M_{D_i}$ and $M^{\perp}_{D_1}$ are, respectively, the nonperturbative functions of the unpolarized and of the polarizing FFs, and $g_K$ is the nonperturbative function of the Collins-Soper Kernel. All other quantities appearing in the above equations, necessary to properly separate the perturbative from the nonperturbative region, are defined and discussed in detail in Ref.~\cite{DAlesio:2022brl}. See also below.

It is worth recalling that Eqs.~(\ref{B0_full_pert}) and (\ref{B1_full_pert}) are obtained by using the leading term of the operator product expansions (OPEs), for small-$\bm{b}_T$ values, of the TMD distribution functions \cite{DAlesio:2022brl,collins_2011}.
Lastly, $S_{\rm pert}$ is the perturbative Sudakov factor, defined as (see also Appendix~\ref{Apx_A} for more details): 
\begin{equation}
    S_{\rm pert}(b_*;\bar{\mu}_{b})=-\widetilde{K}(b_*;\bar{\mu}_{b}) \ln\frac{Q^2}{\bar{\mu}_{b}^2} - \int^{Q}_{\bar{\mu}_{b}} \frac{d\mu'}{\mu'}\,\bigg[ 2\gamma_D(g(\mu');1) - \gamma_K(g(\mu')) \ln{\frac{Q^2}{\mu'^2}} \bigg] \,.
\label{spert_def}    
\end{equation}

The expression of the transverse polarization for the hadron $h_1$ is defined as:
\begin{equation}
    P^{h_1}_n = \frac{d\sigma^{\uparrow} -d\sigma^{\downarrow}}{d\sigma^{\uparrow} +d\sigma^{\downarrow}} = \frac{d\sigma^{\uparrow} -d\sigma^{\downarrow}}{d\sigma^{\rm unp}} \,,
\label{eq:PT_definition}        
\end{equation}
where $d\sigma^{\uparrow(\downarrow)}$ is the differential cross section, Eq.~(\ref{eq:xsec2h}), for the production of a transversely polarized hadron  along the up(down) direction ($\hat{\bm{n}}$) with respect to the production plane,\footnote{Notice that in such a configuration $\sin(\phi_1-\phi_{S_1})=-1$.} and $d\sigma^{\rm unp}$ is the unpolarized cross section.

Finally, we can write the $\bm{q}_T$-integrated transverse polarization as the ratio of the two convolutions in $\bm{b}_T$-space~\cite{DAlesio:2022brl}:
 \begin{equation}
    P^{h_1}_n(z_{h_1},z_{h_2}) =  \frac{ \int d^2\bm{q}_T \, F^{\sin(\phi_1 - \phi_{S_1})}_{TU} }{ \int d^2\bm{q}_T \, F_{UU}} = \frac{M_{1} \int dq_T\;q_T\;d\phi_1 \, \mathcal{B}_1 \Big[\widetilde{D}^{\perp  (1)}_{1T} \widetilde{\bar{D}}_1\Big]}{ \int dq_T\;q_T\;d\phi_1 \, \mathcal{B}_0 \Big[\widetilde{D}_1 \widetilde{\bar{D}}_1\Big] }\,.
\label{pol_ratio}    
\end{equation}

The integration over the azimuthal angle, $\phi_1 $, is trivial. 
Moreover, since the only terms inside the convolutions depending on $q_T $ are the Bessel functions, we can separately integrate them, obtaining 
\begin{equation}
    \int^{q_{T_{\text{max}}}}_0 dq_T \, q_T J_0(b_T\, q_T) = \frac{q_{T_{\text{max}}}}{b_T}J_1(b_T\, q_{T_{\rm max}})\,, \\
\label{int_qtmax_1}    
\end{equation}
\begin{equation}
    \int^{q_{T_{\text{max}}}}_0 dq_T \, q_T J_1(b_T\, q_T) =\frac{\pi q_{T_{\text{max}}}}{2 b_T}
    \{J_1(b_T\, q_{T_{\text{max}}})\bm{H}_0(b_T\, q_{T_{\rm max}}) - J_0(b_T\, q_{T_{\text{max}}})\bm{H}_1(b_T\, q_{T_{\text{max}}}) \}\,,
\label{int_qtmax_2}    
\end{equation}

\noindent where ${\bm{H}}_{0,1}$ are the Struve functions of order zero and one respectively. Notice that in the above integration we have introduced a maximum value $q_{T_{\text{max}}}$, 
that has to fulfil the condition $q_{T_{\text{max}}} \ll Q$, in order to guarantee the validity of the TMD factorization~\cite{Collins:2016hqq}.

\subsection{Semi-inclusive Deep Inelastic Scattering}

Here, we present the formal expressions for the production of a transversely polarized massive hadron $h_1$ in unpolarized SIDIS processes:
\begin{equation}
    e(l)\,N(P)\to e(l')\, h_1(P_1,S_1) +X  \,,
\end{equation}
where $N$ is an unpolarized nucleon with momentum $P$. 
\begin{figure}[!th]
\centering
\includegraphics[width=11cm]{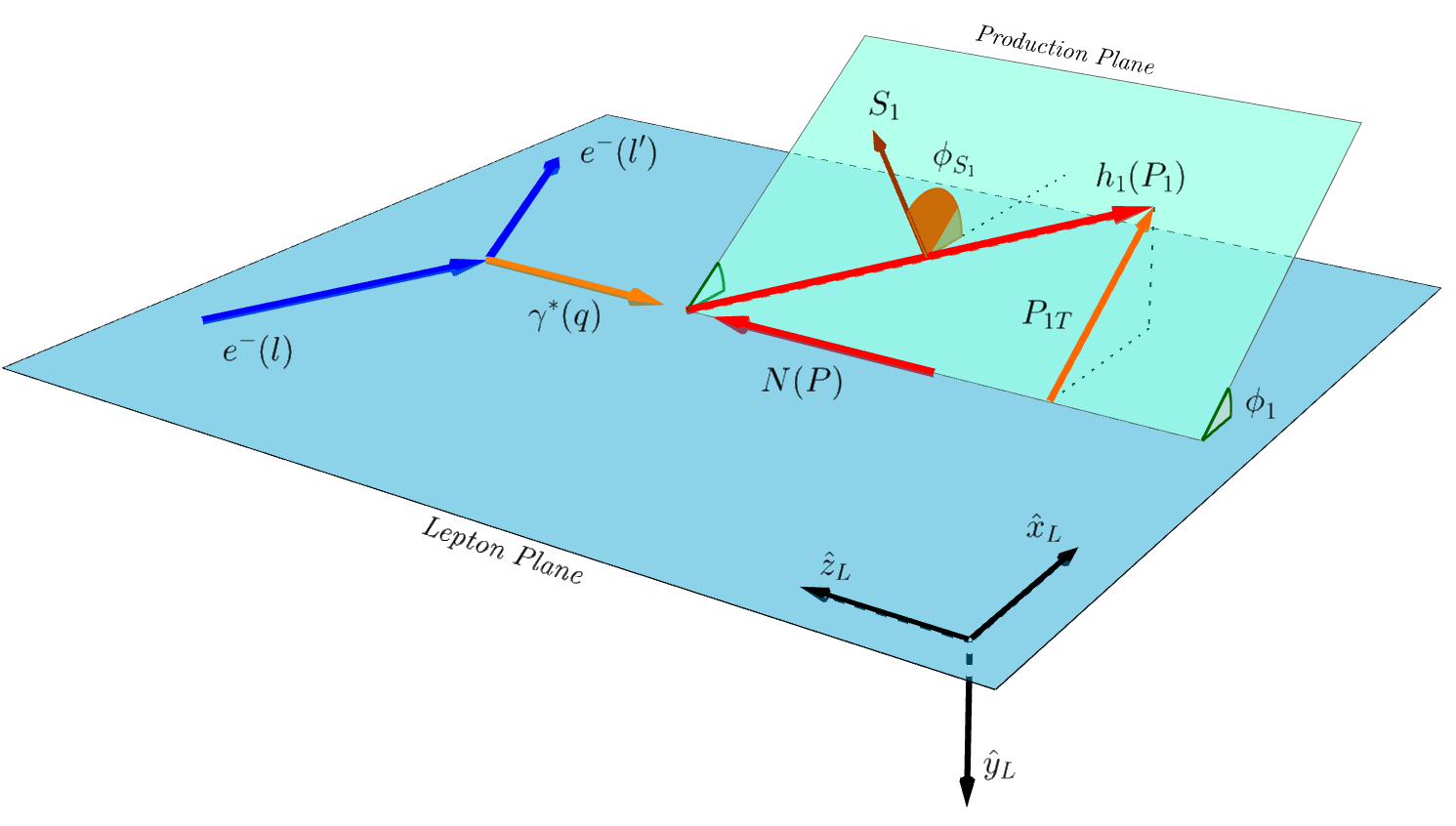}
    \caption{Kinematics for the process $e \,N\to e\, h_1 +X$ in the hadron-frame configuration.}
    \label{fig:kin_sidis}
\end{figure}
In Fig.~\ref{fig:kin_sidis} we show the kinematics of the process in the $\gamma^*N$ c.m.~frame, where the virtual photon, with momentum $q = l- l'$ (virtuality $q^2=-Q^2$), and the nucleon collide along the $\hat{z}_L$ axis, while the hadron $h_1$ moves towards  the negative $\hat{z}_L$ direction with transverse momentum $\bm{P}_{1T}$ with respect to the $\gamma$-$N$ direction. Notice that at variance with the configuration adopted in the ``Trento Conventions'' paper~\cite{Bacchetta:2004jz}, 
the photon moves along $-\hat{z}_L$.
As for the case of the $e^+e^-$ annihilation process, it is more convenient to adopt a frame where the nucleon and the hadron $h_1$ move back to back, along a new $\hat{z}$ axis, and the hadron transverse unbalance is again carried out by the virtual photon. In this frame, the differential cross section, limiting to the terms relevant in the present study, can be written as: 
\begin{equation}
     \frac{d\sigma^{e N \to e h_1(S_1) X}}{ dy dx_B dz_hd^2\bm{q}_{T}} = \sigma^{\rm DIS}_0\bigg[ F_{UU} - |S_{1T}|\sin(\phi_1 - \phi_{S_1}) F^{\sin(\phi_1 - \phi_{S_1})}_{UT} + \dots  \bigg]\,,   
\label{eq:xsecDIS}
\end{equation}
with
\begin{equation}
    x_B=\frac{Q^2}{2 P\cdot q}=x\,, \quad y=\frac{P \cdot q}{ P\cdot l}\,, \quad z_h=\frac{P \cdot P_1}{ P\cdot q}=z = z_p\,,
\end{equation}
where $x = p^+/P^+$ is the light-cone momentum fraction of the nucleon momentum carried by the parton with momentum $p$, and $z$ is the light-cone momentum fraction, defined in Eq.~(\ref{eq:light_con_def}), for the final-state hadron. Notice that the last equalities are exactly true when neglecting the nucleon and the hadron masses, together with terms of order $\mathcal {O} (\bm{k}_\perp^2/Q^2)$.
%
It can be shown that if we keep the final hadron mass (relevant in some kinematical regions)\footnote{The nucleon mass can be safely neglected in our study.} we have
\begin{equation}
    z_p \simeq z_h \Bigg( 1 - \frac{M^2_1}{z^2_h Q^2} \frac{x_B}{1-x_B} \Bigg) \,.
\end{equation}

Another set of invariants adopted in SIDIS, useful from the phenomenological point of view, are the following:
\begin{equation}
         s = (P +l)^2\,, \quad Q^2 =- q^2= x_{B}ys \,, \quad (P + q)^2 = W^2 = \frac{1 - x_B}{x_B}Q^2 \,,
\end{equation}
where $s$ is the total c.m.~energy squared  
and $W$ is the c.m.~energy of the photon-nucleon system. 
In the lepton-nucleon c.m.~frame they can be expressed as:
%
\begin{equation}
    s = 4 E_N E_e\,, \quad Q^2= 4 x_B y E_N E_e\,,
\label{eq:sidis_Q_sep}    
\end{equation}
where $E_{N,e}$ are respectively the nucleon and electron beam energy. Lastly, for the elementary cross section, we have~\cite{Bacchetta:2006tn, Anselmino:2011ch}:
\begin{equation}
    \sigma^{\rm DIS}_0 = \frac{2\pi\alpha^2}{Q^2}\frac{1 + (1-y)^2}{y} \,.
\label{sigma0_dis}    
\end{equation}

In Eq.~(\ref{eq:xsecDIS}) the $F$ terms are now convolutions of a TMD-PDF and a TMD-FF, where again the subscripts denote the polarization states of the initial-state nucleon and the final-state hadron. These are defined as follows~\cite{Mulders:1995dh, Boer:1999uu}:
\begin{eqnarray}
F_{UU} & = & z_{p}^2 \,\mathcal{H}^{(\rm DIS)}(Q) \mathcal{F}[f_1 D_1] \,,\\
F^{\sin(\phi_1 - \phi_{S_1})}_{UT} &=& z_{p}^2\, \mathcal{H}^{(\rm DIS)}(Q) \mathcal{F}\bigg[\frac{\hat{\bm{h}}\cdot \bm{k}_T}{M_{1}}f_1 D^{\perp}_{1T}\bigg]\,, 
\label{eq:main_pol_conv_sidis}    
\end{eqnarray}
where $f_1(x,p_{\perp})$ is the TMD unpolarized parton distribution function and $\mathcal{H}^{(\rm DIS)}(Q)$ is the hard scattering part for the massless on-shell process  $e q\to e q $, at the center-of-mass energy $Q$. Once again at LO this last quantity is normalized to one and  will be dropped in the following.

The convolutions can be written in the conjugate $\bm{b}_T$-space as Fourier transforms:
\begin{eqnarray}
        F_{UU} &=& z_{p}^2 \, \mathcal{B}_0 \Big[\widetilde{f}_1 \widetilde{{D}}_1\Big]= z_{p}^2\, \sum_q e^2_q \int \frac{d b_T}{(2 \pi)} \, b_T J_0(b_T\, q_T) \widetilde{f}_1(x,b_T) \widetilde{{D}}_1(z,b_T)\,, 
\label{eq:conv_sidis_UU}
\end{eqnarray}
\begin{eqnarray}
        F_{UT}^{\sin(\phi_1 - \phi_{S_1})} &=& M_1 \, z_{p}^2 \mathcal{B}_1 \Big[ \widetilde{f}_1  \widetilde{D}^{\perp  (1)}_{1T}\Big]\nonumber\\
        &=&M_{1} z_{p}^2 \sum_q e^2_q \int \frac{d b_T}{2 \pi} \,b^2_T J_1(b_T\,q_T)  \widetilde{{f}}_1(x,b_T)\widetilde{D}^{\perp  (1)}_{1T}(z,b_T) \, , 
\label{eq:conv_sidis_UT}        
\end{eqnarray}
and, after solving the CSS evolution equations, they can be expressed in their full form as:
\begin{eqnarray}
    \mathcal{B}_0 \Big[\widetilde{f}_1 \widetilde{{D}}_1\Big] &=& 
    \frac{1}{z^2}
    \sum_q e^2_q\int \frac{d b_T}{(2 \pi)} \, b_T J_0(b_T\, q_T) \, f_{q/N}(x; \bar{\mu}_b) \, d_{h/q}(z; \bar{\mu}_b)\nonumber \\
    &\times& M_{f_1}(b_c(b_T),x) \, M_{D_h}(b_c(b_T),z)
    \>e^{\!\!-g_K(b_c(b_T);b_{\text{max}})\ln{\big(\frac{Q^2 z}{x M_{P} M_{h}}\big)}- S_{\rm pert}(b_*;\bar{\mu}_{b})}\,,
\label{eq:conv_sidis_UU_bt}    
\end{eqnarray}
\begin{eqnarray}
    \mathcal{B}_1 \Big[ \widetilde{f}_1  \widetilde{D}^{\perp  (1)}_{1T}\Big] &=& 
        \frac{1}{z^2}
    \sum_q e^2_q\int \frac{d b_T}{(2 \pi)} \, b^2_T J_1(b_T\, q_T)  \, f_{q/N}(x; \bar{\mu}_b) \,{D}^{\perp  (1)}_{1T,q} (z;\bar{\mu}_b) \nonumber\\
    &\times&  \,  M_{f_1}(b_c(b_T),x) M^{\perp}_{D_1}(b_c(b_T),z)\>e^{\!\!-g_K(b_c(b_T);b_{\text{max}})\ln{\big(\frac{Q^2 z}{x M_{P} M_{h}}\big)}- S_{\rm pert}(b_*;\bar{\mu}_{b})}\,,
\label{eq:conv_sidis_UT_bt}
\end{eqnarray}
where $f_{q/N}$ is the integrated unpolarized parton distribution function, and $M_{f_1}$ is the nonperturbative component of the unpolarized PDF. 
All the remaining terms that appear in Eq.~(\ref{eq:conv_sidis_UU_bt}) and (\ref{eq:conv_sidis_UT_bt}) are the same defined in the previous section.

The operative expression of the transverse polarization can be obtained from Eq.~(\ref{eq:PT_definition}), where now $d\sigma^{\uparrow(\downarrow)}$ is the differential cross section for a transversely polarized hadron along the up(down) $\hat{\bm {n}}$ direction, with respect to the production plane, in Eq.~(\ref{eq:xsecDIS}).
For nucleons, we can directly write the transverse polarization of the final state hadron and the $\bm{q}_T$-integrated one as the ratio of the two convolutions in $\bm{b}_T $-space:
 \begin{equation}
    P^{h_1}_n(x_B,z_h,q_T) =  \frac{  F^{\sin(\phi_1 - \phi_{S_1})}_{UT} }{ F_{UU}} = \frac{M_{1} \int  d\phi_1 \, \mathcal{B}_1 \Big[ \widetilde{f}_1  \widetilde{D}^{\perp  (1)}_{1T}\Big]}{ \int d\phi_1 \,  \mathcal{B}_0 \Big[\widetilde{f}_1 \widetilde{{D}}_1\Big] }\,,
\label{eq:sidis_pol_ratio}    
\end{equation}
\begin{equation}
    P^{h_1}_n(x_B,z_h) =  \frac{ \int d^2\bm{q}_T \, F^{\sin(\phi_1 - \phi_{S_1})}_{UT} }{ \int d^2\bm{q}_T \, F_{UU}} = \frac{M_{1} \int dq_T\;q_T\;d\phi_1 \,  \mathcal{B}_1 \Big[ \widetilde{f}_1  \widetilde{D}^{\perp  (1)}_{1T}\Big]}{ \int dq_T\;q_T\;d\phi_1 \,  \mathcal{B}_0 \Big[\widetilde{f}_1 \widetilde{{D}}_1\Big] }\,.
\label{eq:sidis_pol_ratio_int}    
\end{equation}

To compute the cross section for the scattering off nuclei, we adopt a simple approach taking the incoherent sum of the contribution of every nucleon that composes the nucleus, neglecting nuclear effects. That is, for the scattering off a nucleus with $A$ nucleons and $Z$ protons we use:
\begin{equation}
    d\sigma^{e A \to e h_1(S_1) X} = Z\, d\sigma^{e p \to e h_1(S_1) X} + (A-Z)\,d\sigma^{e n \to e h_1(S_1) X}\,.
\label{eq:sidis_nuclei}    
\end{equation}

\section{Phenomenology}
\label{4_phenomenology}

In this section, after recalling the main results of the analysis of Belle data~\cite{Belle:2018ttu} presented in Ref.~\cite{DAlesio:2022brl}, we will focus more extensively on the role of the charm contribution and of the $SU(2)$ isospin symmetry.
Then we will give predictions for the transverse $\Lambda$ polarization in $e^+e^-$ collisions, for different values of the c.m.~energy. Finally, we will present estimates for the same observable in semi-inclusive deep inelastic scattering processes, for different values of the lepton and nucleon beam energies. 

\subsection{Two-hadron production data fit: charm and $SU(2)$ isospin symmetry} We begin giving the setup for the phenomenological analysis of Belle data. This is mainly based on our previous work \cite{DAlesio:2022brl}. 
Here we consider only the Belle data set for the polarization of $\Lambda$/$\bar{\Lambda}$ hyperons produced in association with a light hadron, $\pi^{\pm}$ or $K^{\pm}$, measured at $\sqrt{s} = 10.58\,\text{GeV}$. The 128 data points are given as a function of $z_{\Lambda}$ and $z_{\pi/K}$, the energy fractions of $\Lambda$/$\bar{\Lambda}$ and $\pi/K$ particles. For the current analysis, we impose a cut on large values of the light-hadron energy fractions, $z_{\pi/K}<0.5$, keeping only 96 data points, as discussed and motivated in Ref.~\cite{DAlesio:2022brl}. We will come back on this point below.

We will use the following expression to parametrize the $z$ dependence of the first transverse moment of the polarizing $\Lambda$ FF, ${D}^{\perp  (1)}_{1T,\, \Lambda/q}$:
\begin{equation}
     {D}^{\perp  (1)}_{1T,\, \Lambda/q}(z;\mu_b)=\mathcal{N}^{\rm p}_q(z)\,
     d_{\Lambda/q}(z;\mu_b)\,,
\label{eq:first_mom1}     
\end{equation}
with, as adopted and motivated in Ref.~\cite{DAlesio:2020wjq},  $q = u,\, d,\, s,\,\bar{u},\, \bar{d},\, \bar{s}$, and where  
$\mathcal{N}^{\rm p}_q(z)$ (the superscript here refers to the polarizing FF)  is parametrized as:
\begin{equation}
    \mathcal{N}^{\rm p}_q(z) = N_q z^{a_q}(1-z)^{b_q}\frac{(a_q +b_q )^{(a_q +b_q )}}{a_q^{a_q}b_q^{b_q}}\,.
\label{eq:first_mom2}    
\end{equation}
In Eq.~(\ref{eq:first_mom1}), $d_{\Lambda/q}$ is the collinear unpolarized $\Lambda$ fragmentation function for which we employ  
the AKK08 set~\cite{Albino:2008fy}.
This parametrization is given for $\Lambda + \bar{\Lambda}$ and adopts the longitudinal momentum fraction, $z_p$, as scaling variable. In order to separate the two contributions we assume
\begin{equation}
    d_{\bar{\Lambda}/q}(z_p) = d_{\Lambda/\bar{q}}(z_p) = (1- z_p)\, d_{\Lambda/q}(z_p) \,.
\end{equation}
This is a common way to take into account the expected difference between the quark and antiquark FF with a suppressed sea at large $z_p$ as compared to the valence component. Other similar choices have a very little impact on the fit. 

Concerning the nonperturbative function $M^{\bot}_{D, \Lambda}$ we employ the Gaussian model:
\begin{equation}
M^{\bot}_{D, \Lambda}(b_T,z) = \exp{\bigg(-\frac{\langle p_\perp^2 \rangle_\text{p} b^2_T}{4 z^2_p}\bigg)}\,,
\label{eq:pwrlw_polFF}
\end{equation}
where $\langle p_\perp^2 \rangle_{\text{p}} $ is the Gaussian width, a free parameter that we extract from the fit.
Regarding the collinear FFs of the unpolarized light hadrons, $\pi$ and $K$, we adopt the DSS07 set~\cite{deFlorian:2007aj}, while for $M_D$ we consider the PV17 model \cite{Bacchetta:2017gcc}: 
\begin{equation}
    M_D(b_T,z) =\frac{g_3\, e^{-b^2_T\frac{g_3}{4z^2}} +\frac{\lambda_F}{z^2}g^2_4(1 -g_4\frac{b_T^2}{4z^2} )e^{-b^2_T\frac{g_4}{4z^2}} }{g_3 + \frac{\lambda_F}{z^2}g^2_4 }\,,
\label{PV17_md}    
\end{equation}
where
\begin{equation}
    g_{3,4} = N_{3,4}\frac{(z^{\beta}+\delta)(1-z)^{\gamma} }{(\hat{z}^{\beta}+\delta)(1-\hat{z})^{\gamma}} \quad %
\end{equation}
\begin{equation}
\begin{split}
    \hat{z}&= 0.5\,; \quad N_3= 0.21 \,\text{GeV}^2\,; \quad N_4 = 0.13 \,\text{GeV}^2\,;\\
    \beta &= 1.65\,; \quad \delta = 2.28\,; \quad \gamma=0.14\,; \quad \lambda_F= 5.50 \,\text{GeV}^{-2}\,.
\end{split}
\end{equation}
For the $g_K$ function, we use the one extracted in Ref.~\cite{Bacchetta:2017gcc}:
\begin{equation}
        g_K(b_T;b_{\text{max}}) = \frac{g_2  b^2_T}{2}\,; \quad g_2 = 0.13\,\text{GeV}^2  \,.
\label{eq:all_gk}    
\end{equation}
For what concerns the $\Lambda$ unpolarized FF, for $M_D$ we use  a Power-Law model, see Refs.~\cite{DAlesio:2022brl, Boglione:2017jlh,Boglione:2020auc,Boglione:2022nzq}:
\begin{equation}
    M_D(b_T,z,p,m) = \frac{2^{2-p}}{\Gamma(p-1)}\,(b_T m/z_p)^{p-1}{K}_{p-1}(b_T m/z_p)\,,
\label{eq:pwrlw_mod_np}    
\end{equation}
with $p=2$ and $m=1\,\text{GeV}$.
Notice that in the above equations all conversions among the different scaling variables ($z,z_p,z_h$) involved are properly taken into account.

In Eqs.~(\ref{B0_full_pert}), (\ref{B1_full_pert}),  (\ref{eq:conv_sidis_UU_bt}) and (\ref{eq:conv_sidis_UT_bt}) we use the following definition for the $\bar{\mu}_{b}$ variable:
\begin{equation}
    \bar{\mu}_{b} = \frac{C_1}{b_*(b_T)}\, ,
\end{equation}
where $C_1 = 2e^{-\gamma_E}$ (with $\gamma_E$ being the Euler-Mascheroni constant), and the $b_*$ prescription of Ref.~\cite{Bacchetta:2017gcc}:
\begin{equation}
  b_* \equiv b_*(b_T;b_{\rm min},b_{\rm max}) = b_{\rm max}\bigg(\frac{1-e^{-b_T^4/b^4_{\rm max}}}{1-e^{-b_T^4/b^4_{\rm min}}} \bigg)^{1/4}\,.
\end{equation}
Moreover, we adopt
\begin{equation}
    b_c(b_T) = \sqrt{b^2_T + b^2_{\text{min}}}\,,
\end{equation}
with $b_{\text{min}} = 2e^{-\gamma_E}/Q$ and $b_{\text{max}} = 0.6\,\text{GeV}^{-1}$ where, in this analysis, $Q = 10.58 \,\text{GeV}$. %

Since our present goal is a phenomenological analysis at NLL accuracy, for the perturbative Sudakov factor in Eq.~(\ref{spert_def}) we use $\alpha_s$ at LO, the anomalous dimension $\gamma_K$ at the second order, and the Collins-Soper kernel $\Tilde{K}$ and $\gamma_D$ at the first order (see Appendix~\ref{Apx_A} for the explicit expressions of the Sudakov factor and of the anomalous dimensions). A complete N$^2$LL extraction could be achieved only by adopting the coefficient functions, in the OPE, at the next order.

Lastly, for the integration in Eqs.~(\ref{int_qtmax_1}) and~(\ref{int_qtmax_2}), we use $q_{T_{\text{max}}} = 0.27\,Q $. This specific value is chosen on the basis of the results obtained in Ref.~\cite{DAlesio:2022brl} (Fig.~7), where, as shown for this particular choice of nonperturbative functions, the $\chi^2_{\rm dof}$ reaches its minimum.

Concerning the phenomenological analysis and the extraction of the polarizing FFs from Belle data, we consider three different scenarios, exploiting the role of the charm quark contribution and the $SU(2)$ isospin symmetry: 
\begin{enumerate}
    \item Scenario 1. Here we do not include the charm contribution in the unpolarized cross section and we do not impose the $SU(2)$ isospin symmetry. We then extract different $\Lambda$ pFFs for the $u,d, s$ quarks and a single pFF for the sea ($\bar{u}=\bar{d}=\bar{s}$) antiquarks. As discussed in our prevous analysis~\cite{DAlesio:2022brl}, the optimal choice turns out to be an eight-parameter fit: $N_u, N_d, N_s, N_{\rm sea}, a_s, b_u, b_{\rm sea}$ and $\langle p_\perp^2\rangle_{\rm p}$.
    \item Scenario 2. We include the charm contribution in the unpolarized cross section, but we still do not impose the $SU(2)$ isospin symmetry. 
    We continue to extract different $\Lambda$ pFFs for the $u,d, s$ quarks and a single pFF for the sea ($\bar{u}=\bar{d}=\bar{s}$) antiquarks, as in the first scenario. Here, we need to include an extra parameter resulting in a nine-parameter fit: $N_u, N_d, N_s, N_{\rm sea}, a_d, a_s, b_u, b_{\rm sea}$ and $\langle p_\perp^2\rangle_{\rm p}$. 
    \item Scenario 3. We include the charm contribution in the unpolarized cross section and impose $SU(2)$ isospin symmetry for the $u$, $d$ quark pFFs\ , 
    while still adopting different pFFs for the $s$ and $\bar{s}$ quarks. Notice that the AKK08 FF set allows for a slight violation of the $SU(2)$ symmetry; therefore, even imposing $\mathcal{N}_u^{\rm p} = \mathcal{N}_d^{\rm p}$ and $\mathcal{N}_{\bar u}^{\rm p} = \mathcal{N}_{\bar d}^{\rm p}$, the extracted pFFs will be still slightly different, see below. In such a case the nine free parameters are: $N_{u,d}, N_{\bar u,\bar d}, N_s, N_{\bar s}, a_{u,d}, a_s, b_{u,d}, b_{\bar s}$ and $\langle p_\perp^2\rangle_{\rm p}$. Notice that the inclusion of a further parameter for the sea pFFs, namely $b_{\bar u,\bar d}$, does not improve the quality of the fit.
\end{enumerate}
As already discussed in our previous analyses, the imposition of the $SU(2)$ symmetry alone within a three-flavor scheme would lead to a very poor quality of the fit. 

\begin{figure}[!bt]
\centering
{\includegraphics[trim =  0 50 0 140,clip,width=7.5cm]{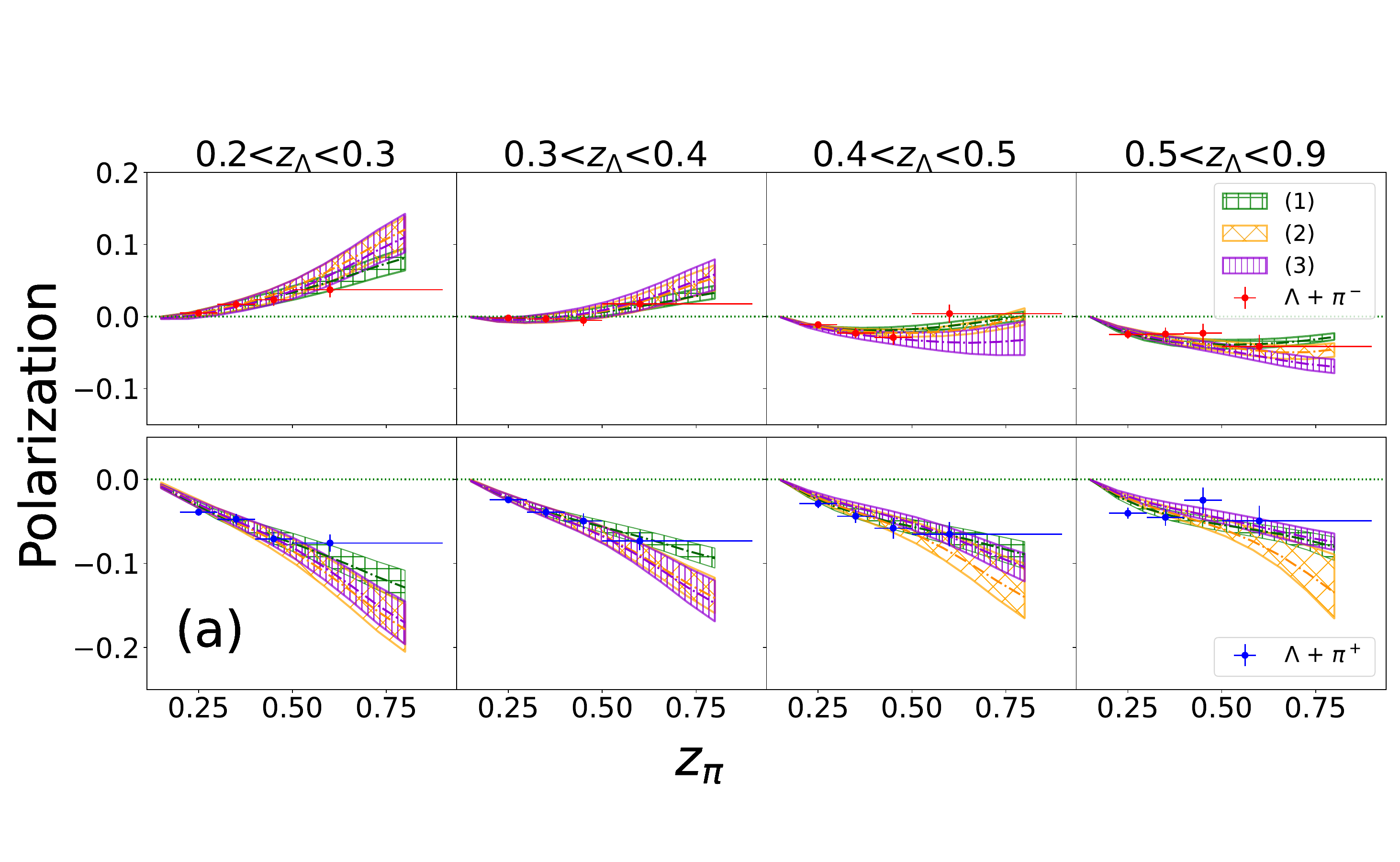}}
{\includegraphics[trim =  0 50 0 140,clip,width=7.5cm]{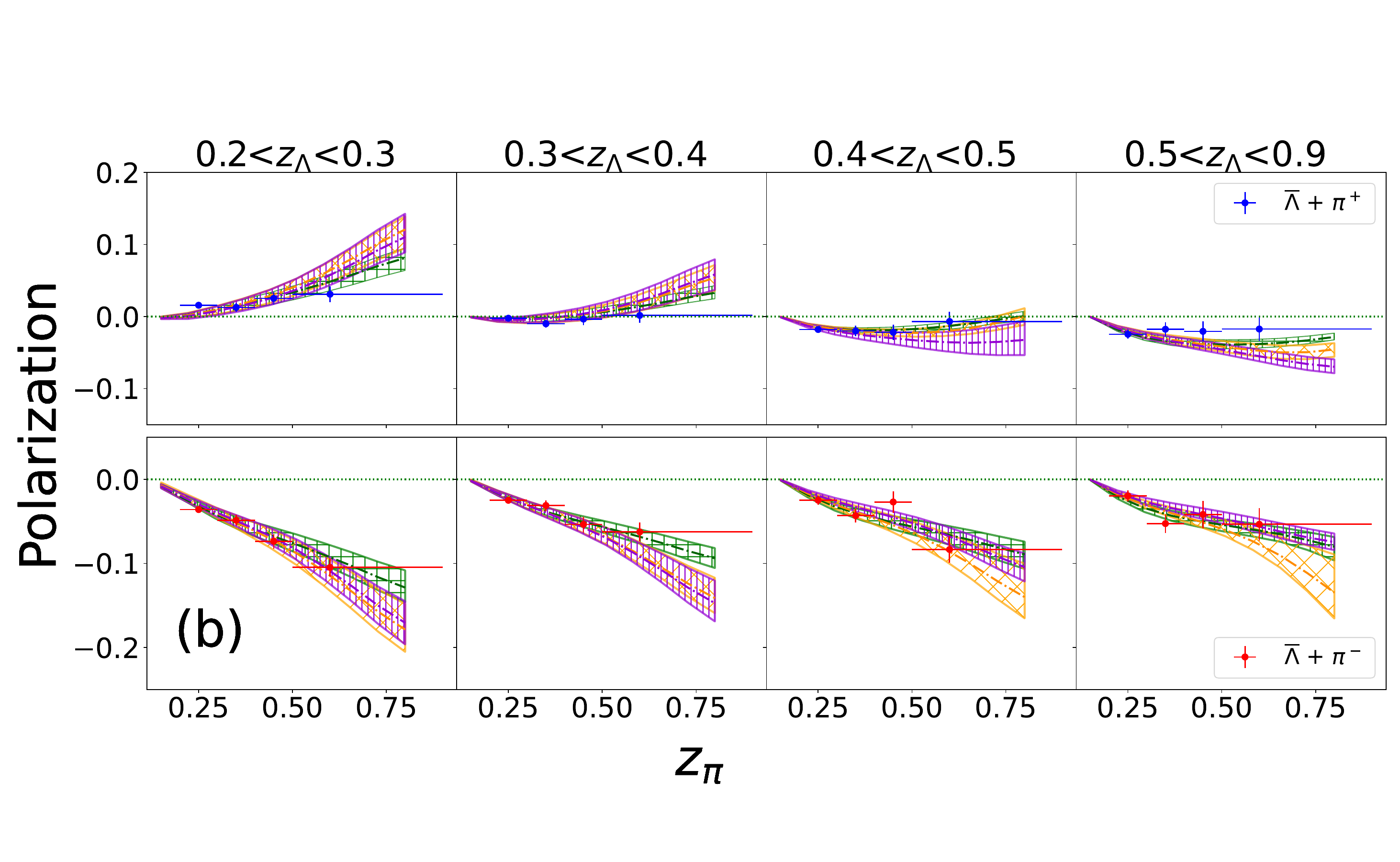}}
\newline
{\includegraphics[trim =  0 50 0 140,clip,width=7.5cm]{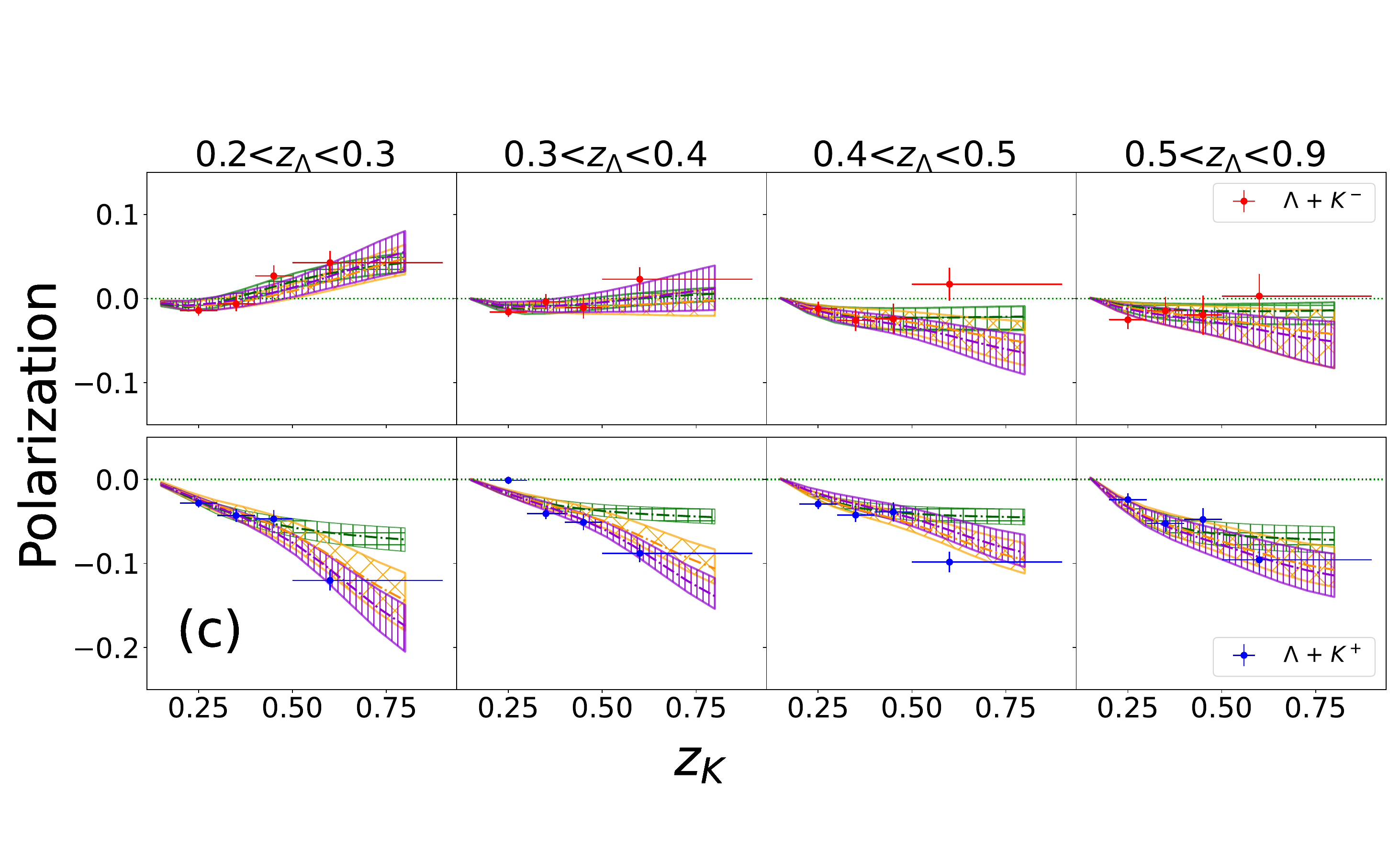}}
{\includegraphics[trim =  0 50 0 140,clip,width=7.5cm]{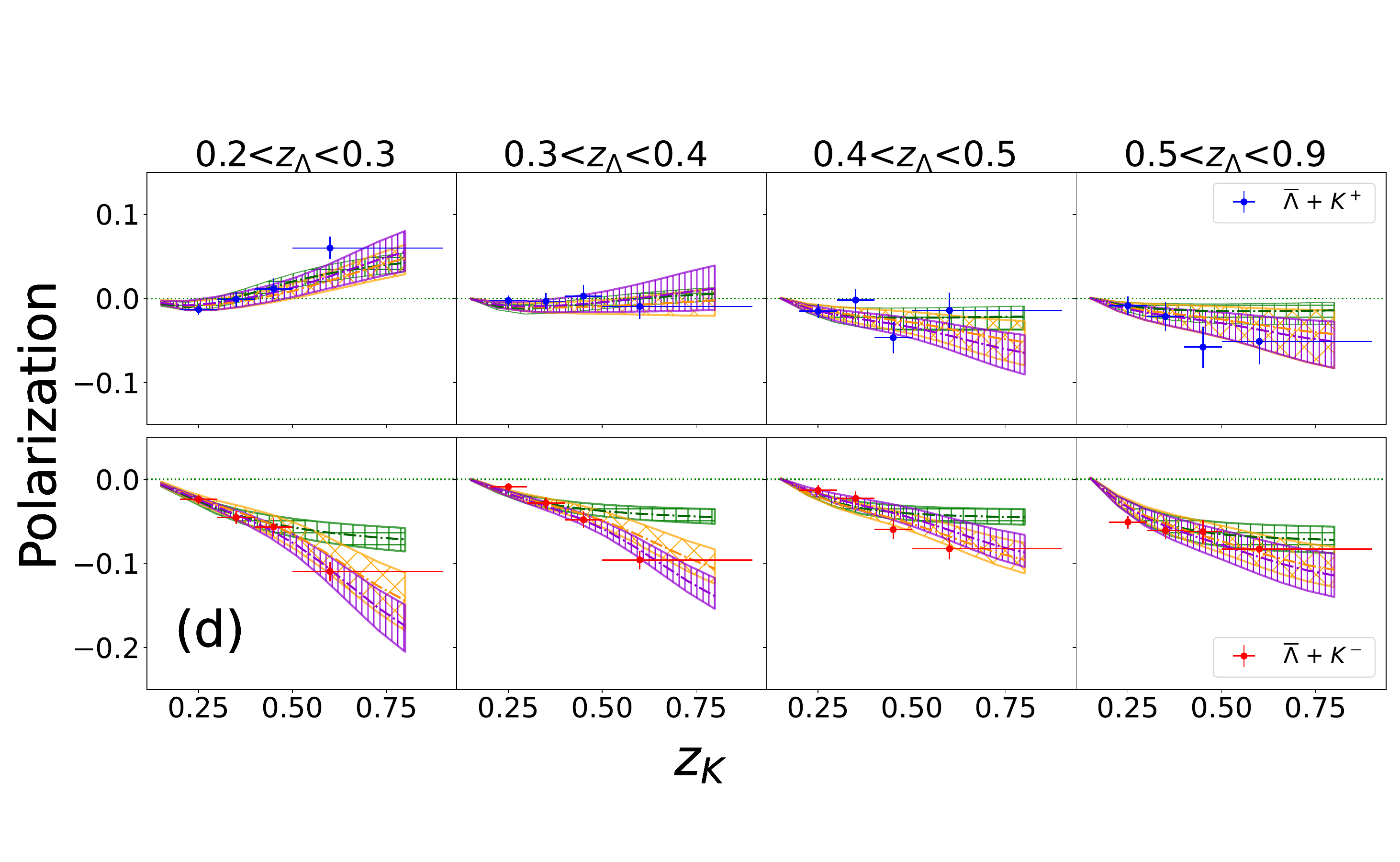}}
\caption{Best-fit estimates of the transverse polarization for $\Lambda$, $\bar{\Lambda}$ in $e^+e^-\to \Lambda (\bar{\Lambda}) h +X $, for $\Lambda \pi^{\pm}$ (a), $\bar{\Lambda} \pi^{\pm}$ (b), $\Lambda K^{\pm}$ (c), $\bar{\Lambda} K^{\pm}$ (d), as a function of $z_h$ ($h= \pi, K$) for different $z_{\Lambda}$ bins, adopting the three different scenarios: (1) green, (2) orange and (3) violet bands. Data are from Belle~\cite{Belle:2018ttu}. The statistical uncertainty bands, at $2\sigma$ level, are also shown. Data for $z_{\pi, K} > 0.5$ are not included in the fit.}
\label{fig:Lh_gauss} 
\end{figure}

The best-fit parameters extracted for the first moment of the pFFs are given in Tab.~\ref{tab:parameters_12}, together with the $\chi^2_{\rm dof}$s for each scenario, while in Fig.~\ref{fig:Lh_gauss} we show the corresponding estimates of the transverse $\Lambda$, $\bar{\Lambda}$ polarizations, produced in association with a light-hadron, compared against Belle data~\cite{Belle:2018ttu}.

\begin{table}[!th]
\caption{Best-fit parameter values for the first moment of the polarizing FF and the nonperturbative function employed to fit the double-hadron data set, adopting the three scenarios.}
    \begin{tabular}{|ccc|}
    \hline 
    & \multicolumn{2}{c|}{Scenario}\\
    \hline
    Parameters & (1) & (2) \\ [.7ex]
    \hline
    $N_u$ & $0.144_{-0.076}^{+0.122}$ & $0.178_{-0.096}^{+0.171}$  \\ [.7ex]
    $N_d$ & $-0.140_{-0.057}^{+0.034}$ & $-0.376_{-0.439}^{+0.262}$   \\ [.7ex]
    $N_s$ & $-0.151_{-0.102}^{+0.071}$ & $-0.121_{-0.098}^{+0.053}$ \\ [.7ex]
    $N_{\rm sea}$ & $-0.09_{-0.099}^{+0.054}$ & $-0.127_{-0.129}^{+0.079}$\\ [.7ex]
    $a_d$ &      & $0.774_{-0.773}^{+1.074}$ \\ [.7ex]
    $a_s$ & $2.046_{-0.732}^{+0.967}$ & $0.848_{-0.553}^{+0.913}$ \\ [.7ex]
    $b_u$ & $3.57_{-1.471}^{+2.017}$ & $2.71_{-1.511}^{+2.387}$  \\ [.7ex]
    $b_{\rm sea}$ & $2.606_{-1.596}^{+2.629}$ & $1.59_{-1.294}^{+2.138}$ \\ [.7ex]
    $\langle p_\perp^2 \rangle_{\rm p}$ & $0.097_{-0.034}^{+0.045}$ &$ 0.093_{-0.045}^{+0.054}$ \\ [.7ex]
    \hline
    $\chi^2_{\rm dof}$&1.174 & 1.259 \\[.7ex]
    \hline
\end{tabular}
\begin{tabular}{|cc|}
    \hline 
    & Scenario\\
    \hline
    Parameters & (3)  \\ [.7ex]
    \hline
    $N_{u,d}$ & $0.130_{-0.031}^{+0.036}$   \\ [.7ex]
    $N_{\bar{u},\bar{d}}$ & $-0.174_{-0.048}^{+0.052}$  \\ [.7ex]
    $N_s$ & $-0.263_{-0.183}^{+0.136}$  \\ [.7ex]
    $N_{\bar{s}}$ & $-0.150_{-0.06}^{+0.056}$ \\ [.7ex]
    $a_{u,d}$ & $2.838_{-2.213}^{+4.565}$ \\ [.7ex]
    $a_s$ & $2.164_{-0.903}^{+1.158}$ \\ [.7ex]
    $b_{u,d}$ & $10.58_{-7.004}^{+13.56}$   \\ [.7ex]
    $b_{\bar{s}}$ & $1.545_{-0.818}^{+0.926}$  \\ [.7ex]
    $\langle p_\perp^2 \rangle_p$ & $0.116_{-0.042}^{+0.042}$  \\ [.7ex]
    \hline
    $\chi^2_{\rm dof}$&1.361  \\[.7ex]
    \hline
\end{tabular}
\label{tab:parameters_12}
\end{table}
Some comments are in order here: Comparing the parameter values obtained within the first and second scenario, we see a significant difference in their magnitudes, somehow due to the inclusion of the charm quark contribution in the second one. However, as already shown in our previous works~\cite{DAlesio:2020wjq,DAlesio:2022brl}, in both cases only the up pFF is positive, while the remaining pFFs are all negative. On the contrary, within the third scenario, we observe that both the up and down pFF are positive, having an opposite sign with respect to the anti-up and anti-down pFFs. The strange and anti-strange pFFs come out still negative. The main point in this comparison is that if we allow for different normalization factors for the up and down pFFs (Sc.~1 and Sc.~2), they come out opposite in sign, leading to a strong violation of the $SU(2)$ symmetry. And this happens even if we allow for independent normalization factors for the sea contributions. In other words, only imposing $N_d=N_u$ we can restore, at least approximately with this set of unpolarized FFs, the symmetry.

\begin{center}
\begin{figure}[!th]
\includegraphics[trim =  70 50 250 0,clip,width=7cm]{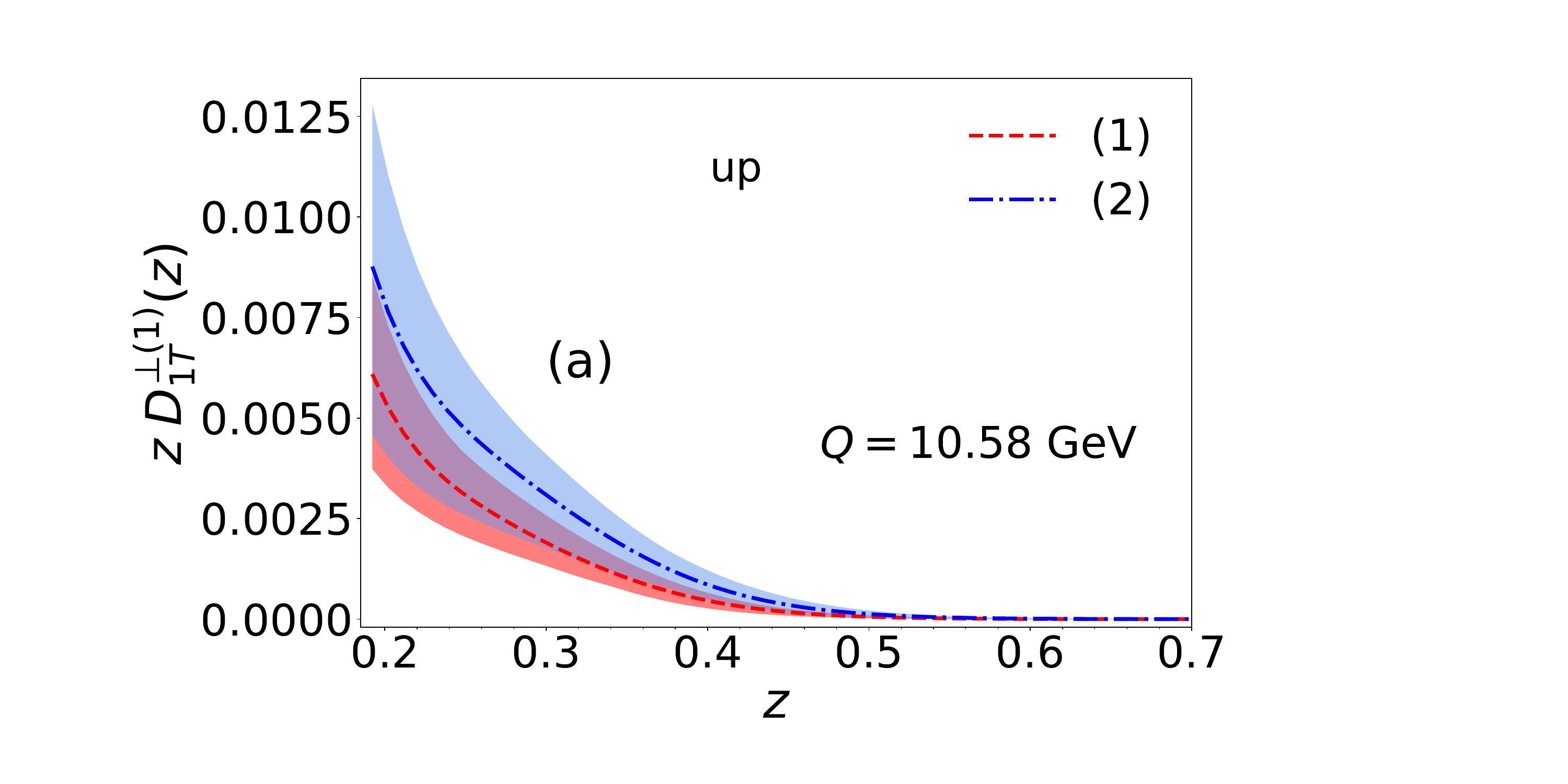}
\includegraphics[trim =  70 50 250 0,clip,width=7cm]{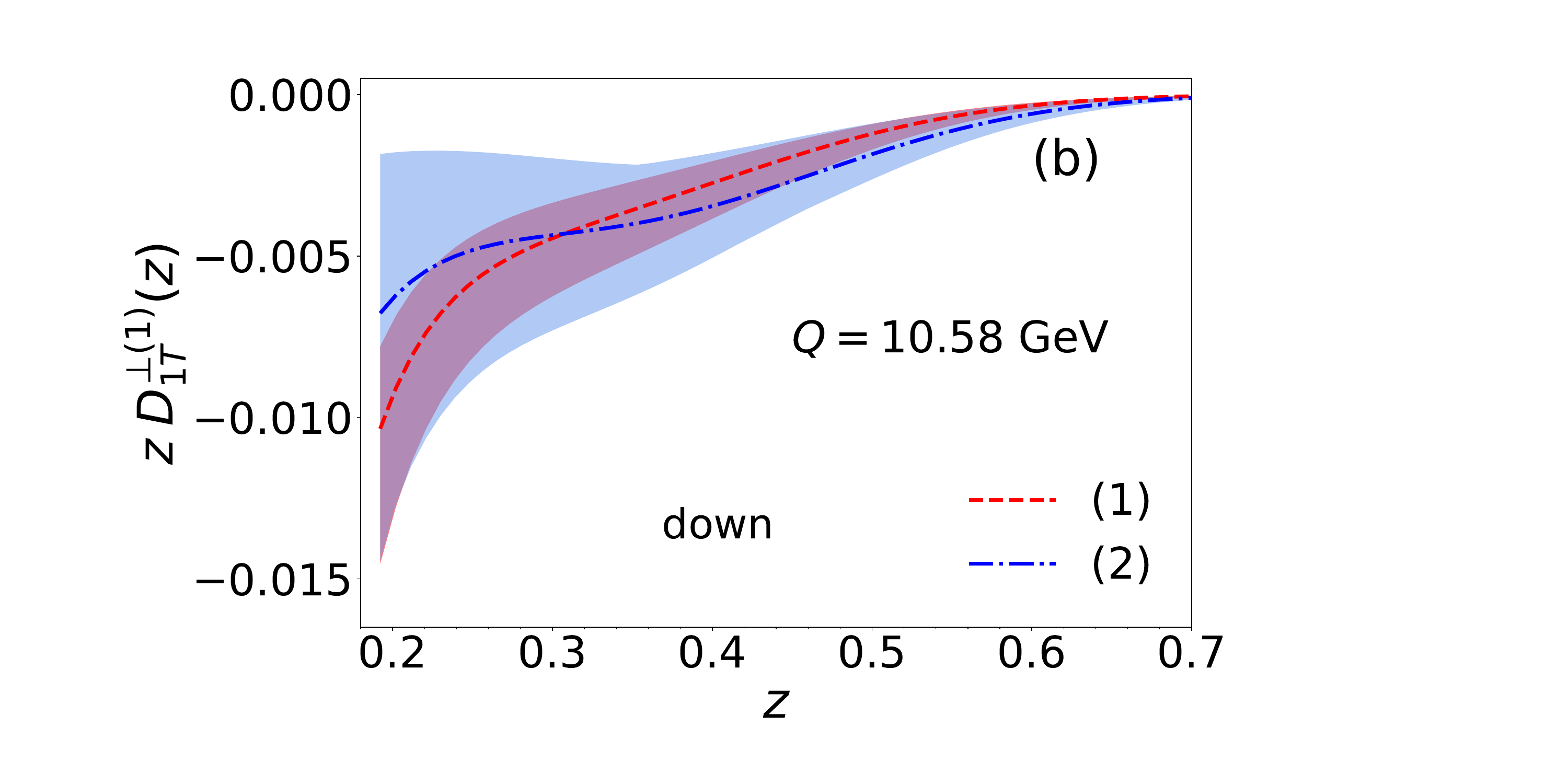}\newline
\includegraphics[trim =  70 50 250 0,clip,width=7cm]{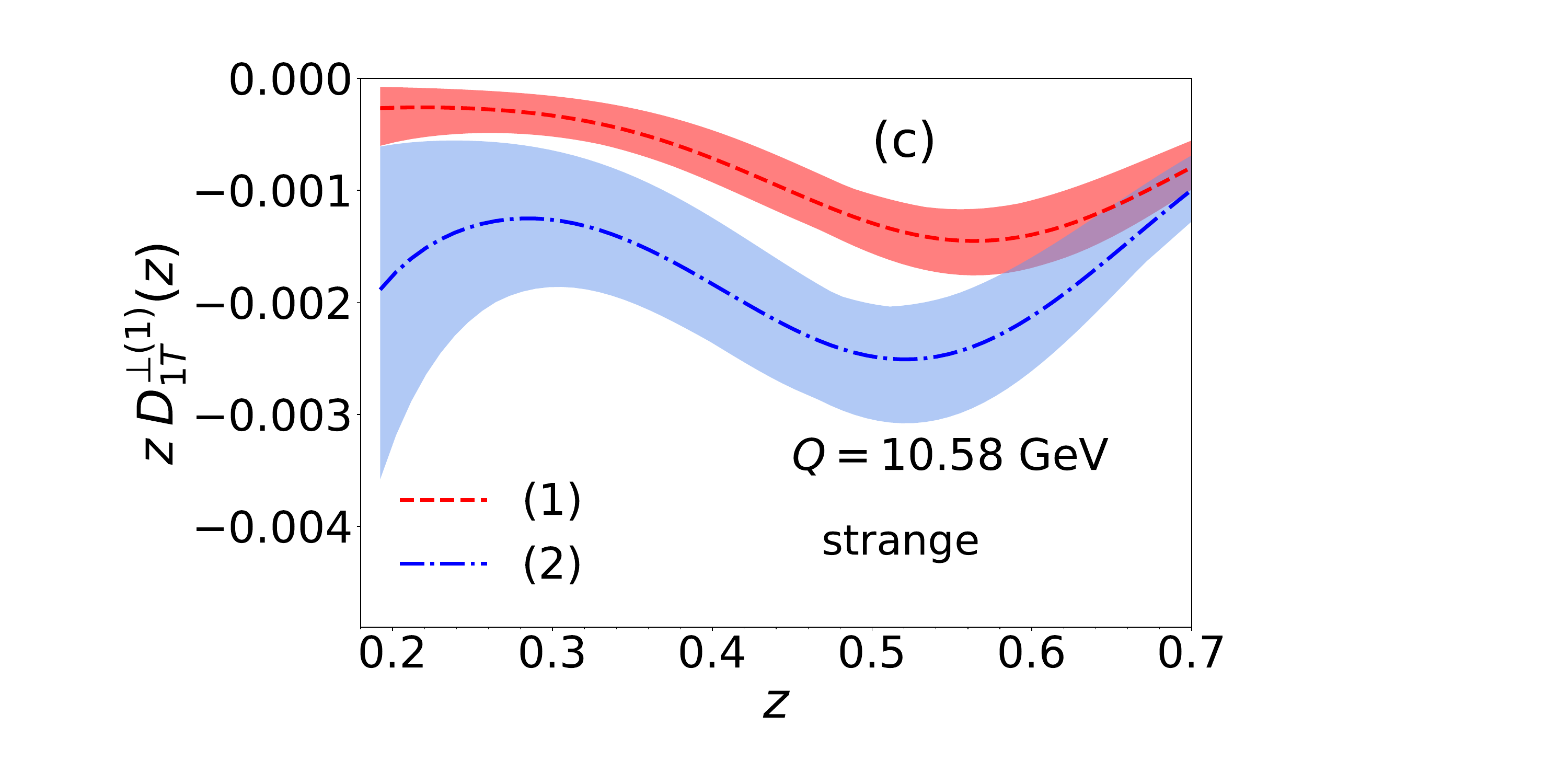}
\includegraphics[trim = 70 50 250 0,clip,width=7cm]{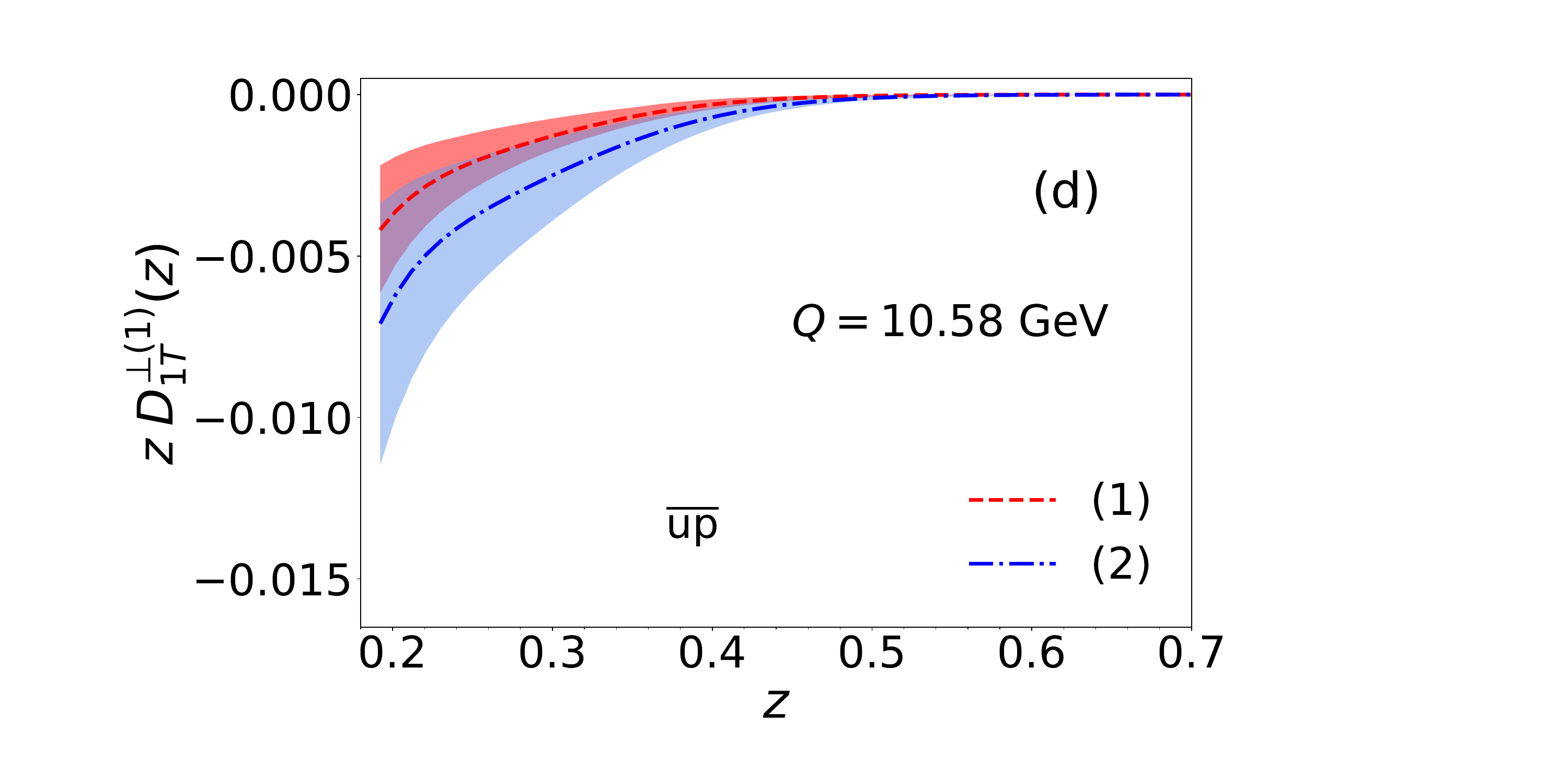}
\caption{First moments of the polarizing FFs, for the up (a), down (b), strange (c) and sea (d) quarks, as obtained from the  fit within the first (red dashed lines) and the second (blue dot-dashed lines) scenarios. The corresponding statistical uncertainty bands, at $2\sigma$ level, are also shown. }
\label{fig:first_mom12} 
\end{figure}
\end{center}

\begin{center}
\begin{figure}[!thb]
\includegraphics[trim =  70 50 250 0,clip,width=7cm]{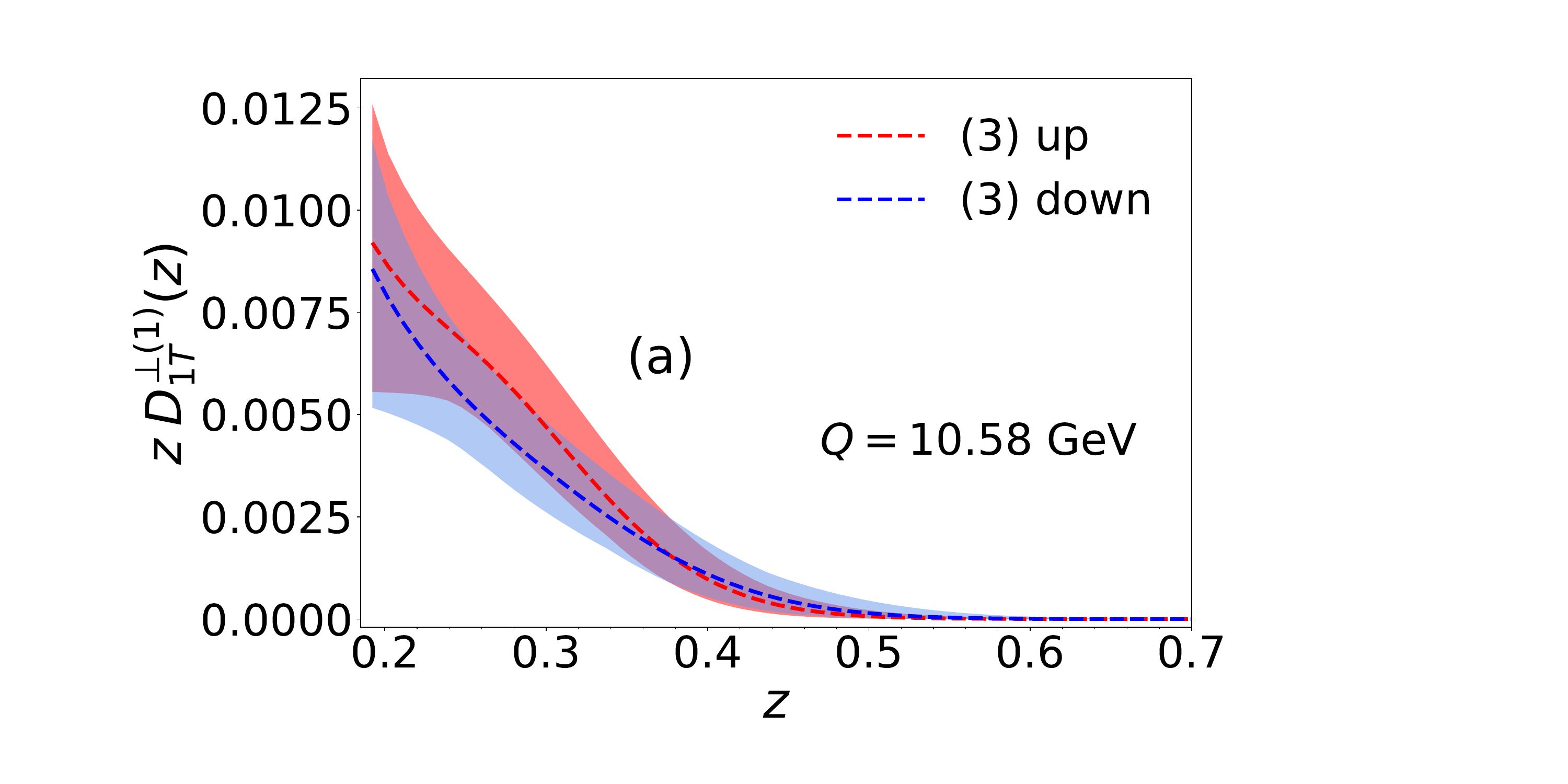}
\includegraphics[trim =  70 50 250 0,clip,width=7cm]{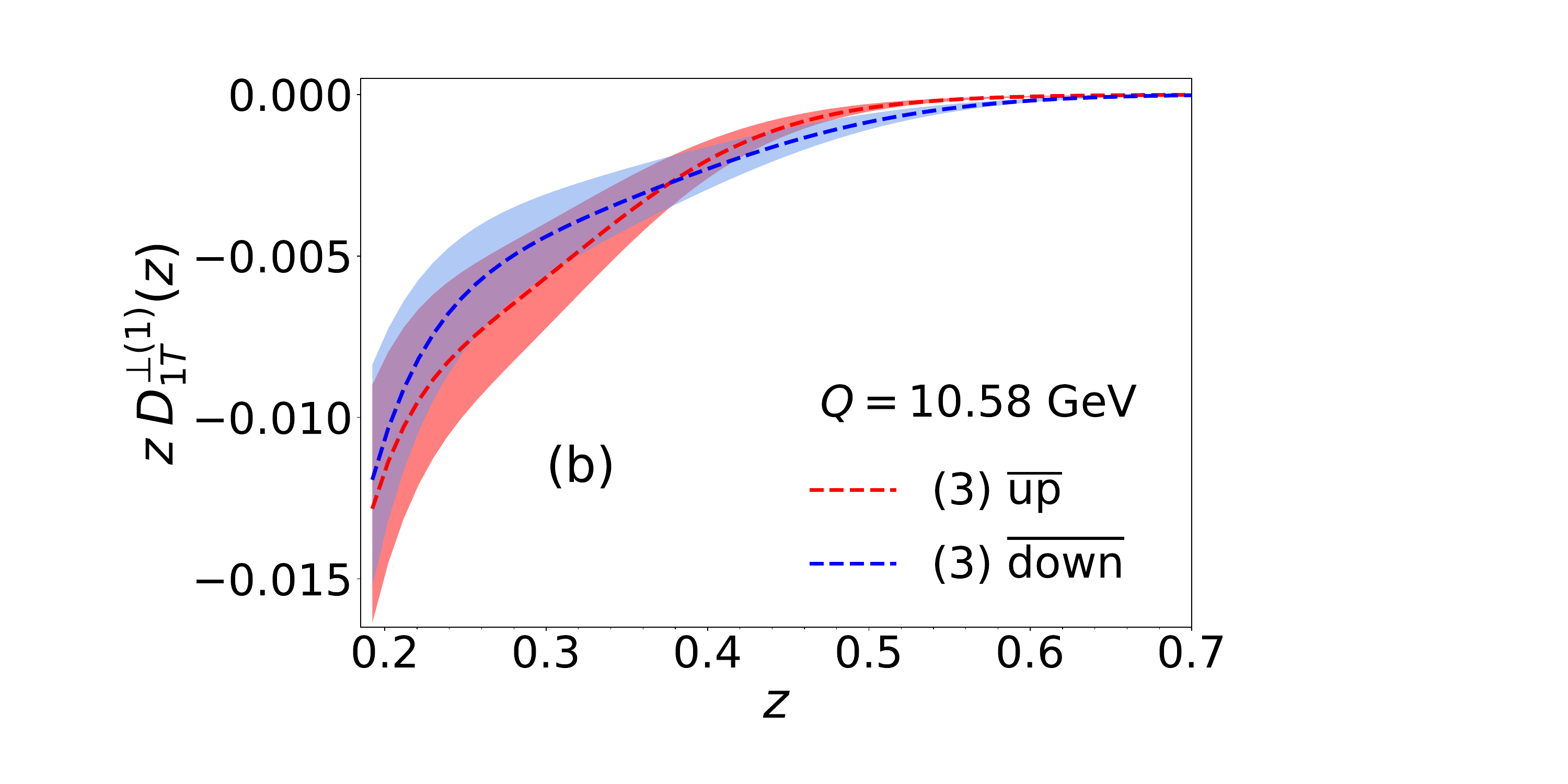}\newline
\includegraphics[trim =  70 50 250 0,clip,width=7cm]{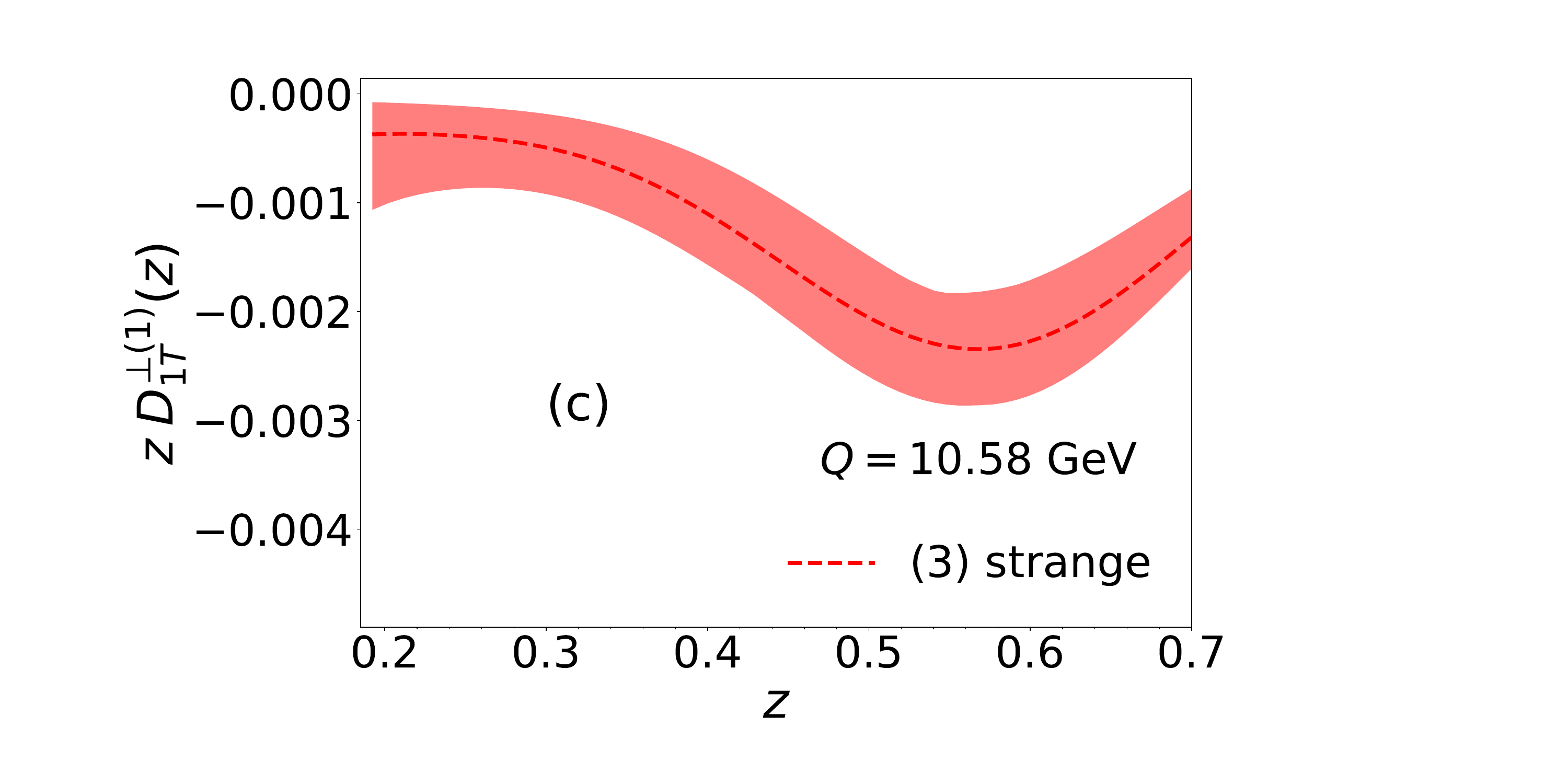}
\includegraphics[trim = 70 50 250 0,clip,width=7cm]{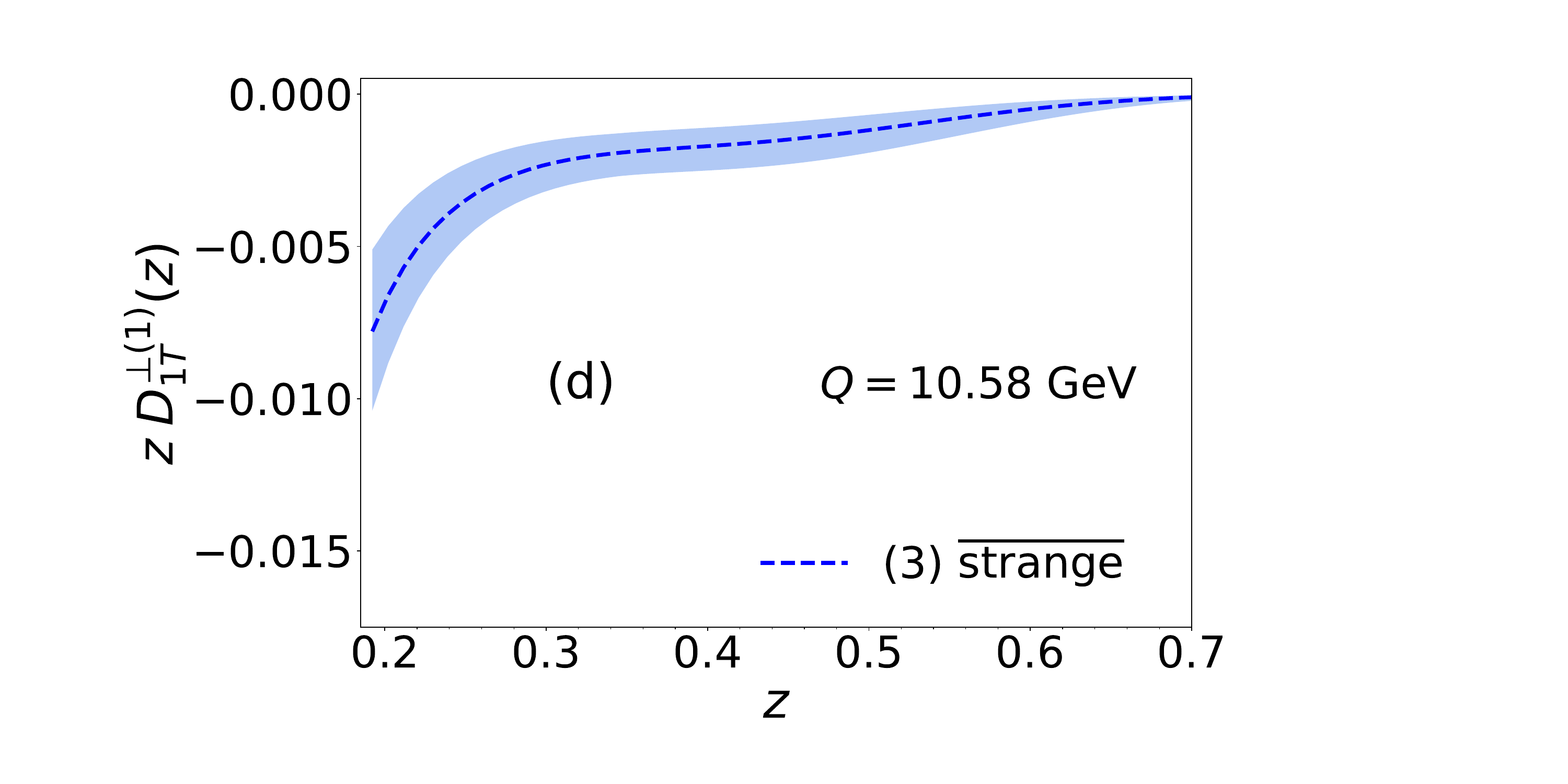}
\caption{First moments of the polarizing FFs, for the up/down (a), anti-up/down (b), strange (c) and anti-strange (d) quarks, as obtained from fit within the third scenario. The corresponding statistical uncertainty bands, at $2\sigma$ level, are also shown. }
\label{fig:first_mom3} 
\end{figure}
\end{center}

Despite these differences, within all scenarios we obtain similar sizes for the Gaussian width. 
The first $k_\perp$-moments of the polarizing FFs are shown in Fig.~\ref{fig:first_mom12} for scenarios 1 and 2, and in Fig.~\ref{fig:first_mom3} for scenario 3. 
In scenarios 1 and 2 the first moments are all compatible, at least within the uncertainty bands, with the exception of the strange pFF. When we move to the third scenario the up pFF comes out still compatible with the results in the other scenarios, with the strange pFF somehow in between. The most interesting finding is that since in this scenario the down pFF is positive ($SU(2)$  constrained), the negative sea contributions are larger in size.

It is important to stress here that in the extraction of the first moment of the polarizing FFs we do not impose any positivity bound, that, in principle, could prevent a proper sampling of the parameter space. On the other hand we have checked, \emph{a posteriori}, that  this is fulfilled in all scenarios considered.

Moving to the comparison with data, we can generally say that all three scenarios are able to describe reasonably, or even quite, well the $\Lambda\pi^{\pm}$, $\bar{\Lambda}\pi^{\pm}$, $\Lambda K^{-}$ and $\bar{\Lambda} K^{+}$ polarization data. However, as already pointed out in our first works, where we did~\cite{DAlesio:2022brl} or did not~\cite{DAlesio:2021dcx} employ the full TMD machinery, within scenario 1 one cannot describe, at variance with the $\Lambda\pi$ case, the $\Lambda K^+$ and $\bar\Lambda K^-$ data with $z_{K}>0.5$. 


Quite interestingly, when we include the charm contribution, imposing or not the $SU(2)$ isospin symmetry, we can still obtain similar good fits with a simultaneous very good description of these data points (even if not included in the analysis), see Figs.~\ref{fig:Lh_gauss}c and \ref{fig:Lh_gauss}d, lower panels. 
%
This result, focusing on $\Lambda K^+$ for simplicity, can be understood as follows: in scenarios~2 and 3 the inclusion of the charm contribution in the denominator, with non negligible charm FFs both for $K^+$'s and $\Lambda$'s, requires larger, in size, pFFs in the fit. 
Moreover, since this extra piece in the denominator happens to be a decreasing function in $z_K$ 
the polarization eventually increases in size with $z_K$.

From the present study, it is very likely that the inclusion of the charm contribution, at least in the unpolarized cross section, must be considered necessary for the analysis of Belle data. Several attempts to include this contribution also in the numerator of the transverse polarization (that is parametrizing also pFFs for charm quarks) have been carried out but no significant improvement on the $\chi^2_{\rm dof}$ value or in the description of data has been found. 

Similar conclusions, even if on a more qualitative ground since they do not provide any $\chi^2$ value and any uncertainty band, have been obtained in Ref.~\cite{Chen:2021hdn}. Here, by including the charm contribution, also for the polarizing FFs (resulting in a 20-parameter fit) they show that Belle data can be described reasonably well even without any isospin symmetry violation.  

In this respect, we agree that the issue of $SU(2)$ symmetry has to be taken with care and that cannot be solved by analysing only the data on the transverse polarization of $\Lambda/\bar\Lambda$ produced in $e^+e^-$ processes. More experimental information is therefore certainly needed.

\subsection{Predictions for the transverse $\Lambda$ polarization in $e^+e^-$ collisions at different energies}

\begin{figure}[t! ]
{\includegraphics[trim =  0 50 0 100,clip,width=7.5cm]{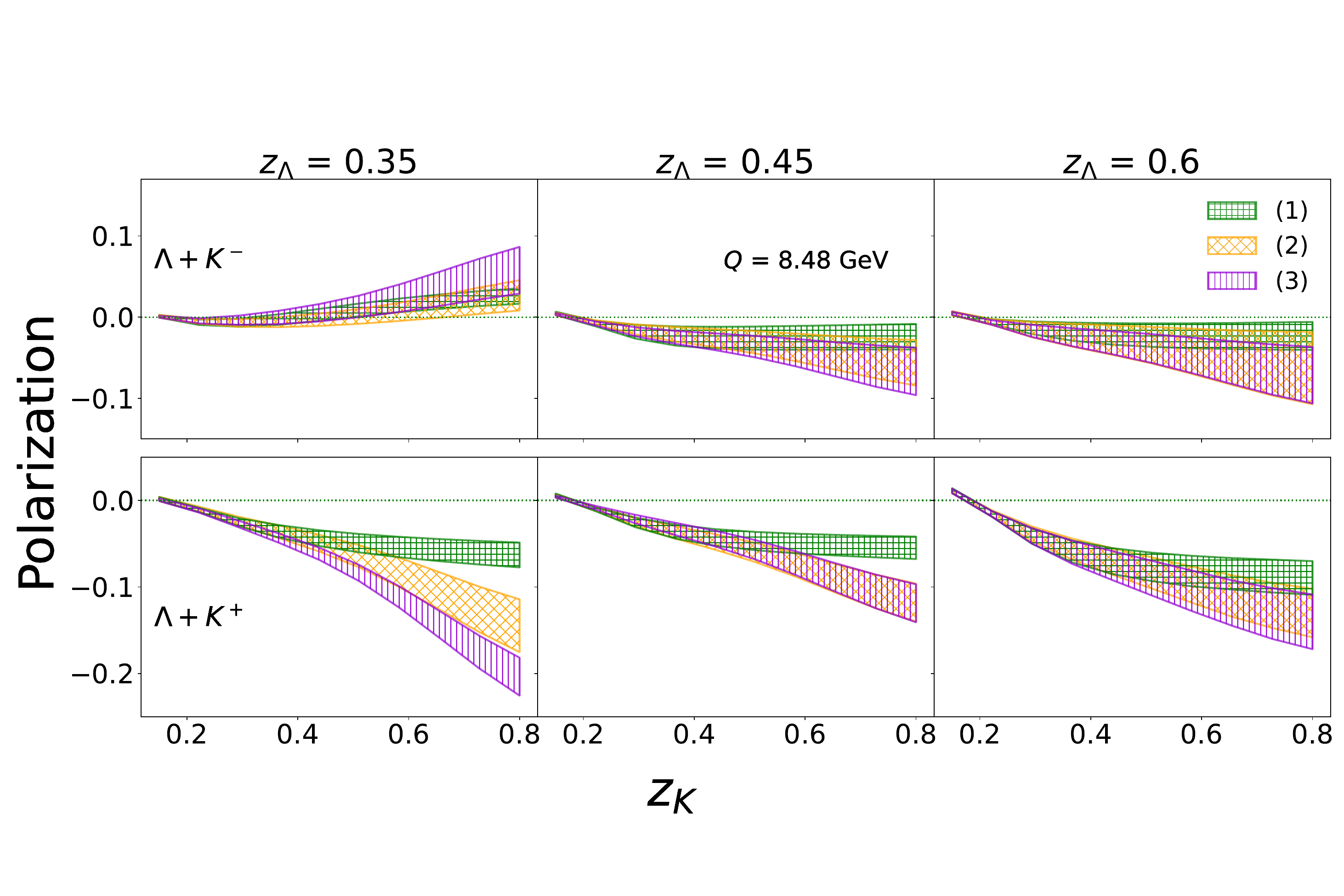}}
{\includegraphics[trim =  0 50 0 100,clip,width=7.5cm]{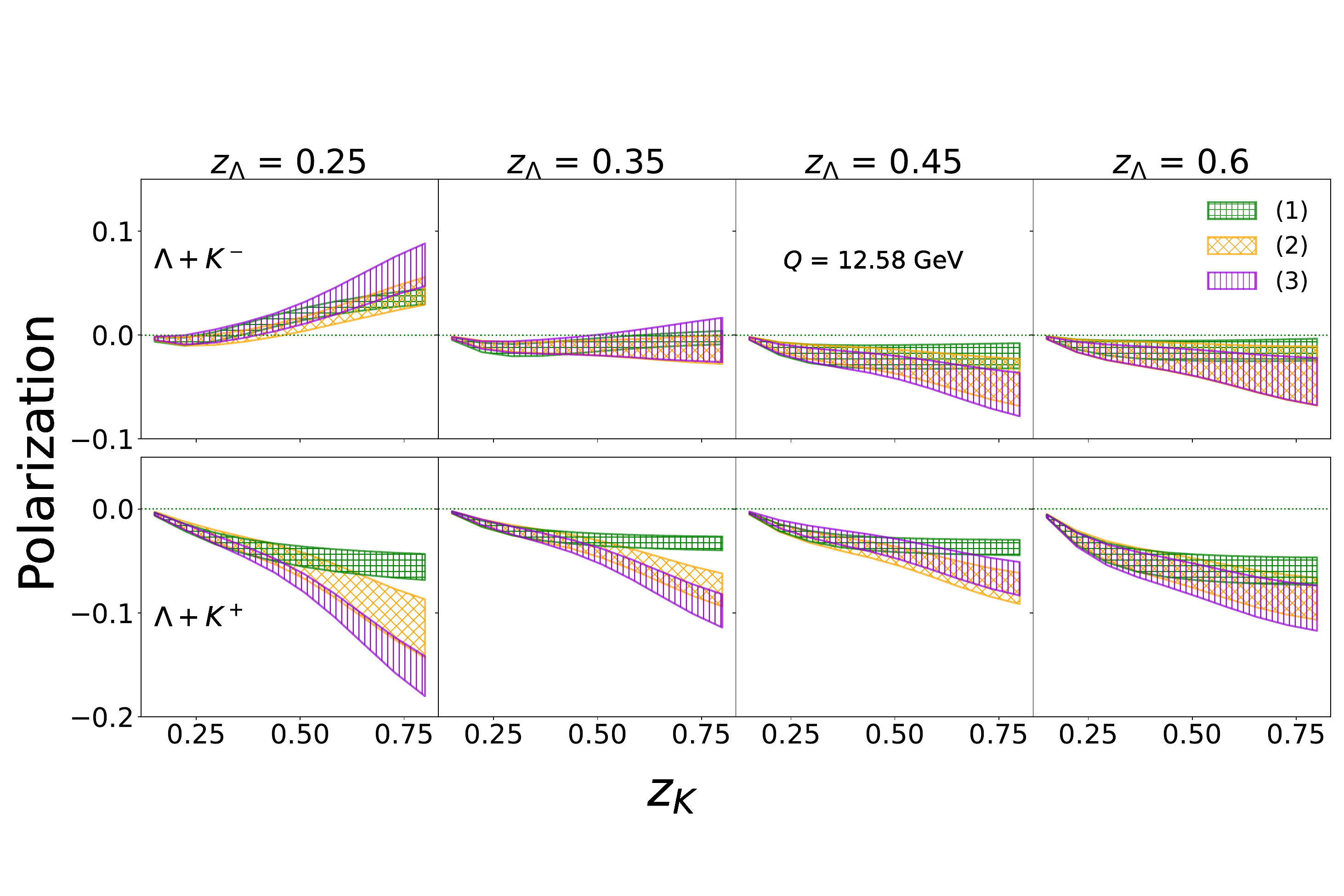}}
\caption{Estimates of the transverse $\Lambda$ polarization in $e^+e^-\to \Lambda K^{\pm} +X $  at $Q = 8.48 \,\text{GeV}$ (left panel), and at $Q = 12.58 \,\text{GeV}$ (right panel), as a function of $z_{K}$ for different $z_{\Lambda}$ bins, for the three different scenarios: (1) green, (2) orange and (3) violet bands. The statistical uncertainty bands, at $2\sigma$ level, are also shown.}
\label{fig:epem_pred} 
\end{figure}

Here we give some predictions at different energies, focusing on $\Lambda$-$K$ production, with the aim to look for possible significant differences among the three scenarios. 
In Fig.~\ref{fig:epem_pred} we show the estimates for the transverse polarization of $ \Lambda$'s produced with $K^{\pm}$ mesons, at different energies, namely, 8.48~GeV (left panel) and 12.58~GeV (right panel). Notice that in the first case we cannot have the $z_{\Lambda}=0.25$ bin for kinematical reasons.
At both energies, only the first $z_\Lambda$ bins show some discrepancies at large $z_K$ values between the predictions obtained within scenarios 2 and 3. On the other hand, for higher values of $z_{\Lambda}$ all predictions become very similar, within the uncertainties. This, also true for high energy values, could prevent the distinction between the two scenarios in future $e^+e^-$ measurements. 

\subsection{Predictions for the transverse $\Lambda$ polarization in SIDIS}

In this section, by using Eqs.~(\ref{eq:conv_sidis_UU_bt}), (\ref{eq:conv_sidis_UT_bt}), (\ref{eq:sidis_pol_ratio}) and (\ref{eq:sidis_pol_ratio_int}), we present the predictions for the transverse $\Lambda$ polarization in unpolarized SIDIS electron-proton and electron-deuterium collisions, for different values of their beam energies, and considering the three previously presented scenarios. 

A similar analysis, showing also estimates obtained from our previous extraction~\cite{DAlesio:2022brl}, has been presented in Ref.~\cite{Chen:2021zrr}. In this paper they consider in detail the transverse $\Lambda$ polarization only, including the charm contribution and imposing $SU(2)$ symmetry (see Ref.~\cite{Chen:2021hdn} and our comments above). In this respect, our predictions are indeed in qualitative agreement with theirs. On the other hand, they do not show any result for the $\bar\Lambda$ case, that, as we will discuss below, represents a much powerful tool to discriminate  different scenarios. It is also worth noticing that a comprehensive phenomenological impact study on the transverse $\Lambda$ polarization at the future EIC, even if limited to scenario 1 and at LO accuracy, has been carried out in  Ref.~\cite{Kang:2021kpt}. In this work they include EIC pseudodata to reweight the parametrization of the polarizing FFs as extracted from Belle $e^+e^-$ data, leading to a significant reduction in the theoretical uncertainties.  

Concerning the present analysis, for $\Lambda$ hyperon, we employ the unpolarized nonperturbative function in Eq.~(\ref{eq:pwrlw_mod_np}), and the polarizing FF first moment and nonperturbative function in Eqs.~(\ref{eq:first_mom1}), (\ref{eq:first_mom2}) and (\ref{eq:pwrlw_polFF}), adopting the parameters given in Tab.~\ref{tab:parameters_12}. As for the proton PDFs, we use for the unpolarized ones the CT14 NNLO  set~\cite{Dulat:2015mca}, and for the nonperturbative function the one extracted in Ref.~\cite{Bacchetta:2017gcc}, which has the following form:
\begin{equation}
    M_{f_1}(b_T,x) = \frac{1}{2 \pi} e^{-g_{1} \frac{b_T^2}{4}} \bigg(1-\frac{\lambda g^2_{1}}{1+ \lambda g_{1}} \frac{b_T^2}{4}\bigg)\,,
\end{equation}
where
\begin{equation}
    g_{1} = N_{1}\frac{(1-x)^{\alpha}x^{\sigma} }{(1-\hat{x})^{\alpha}\hat{x}^{\sigma}} \quad %
\end{equation}
\begin{equation}
\begin{split}
    \hat{x}&= 0.1 \quad N_1= 0.28 \,\text{GeV}^2 \\
    \alpha &= 2.95 \quad \sigma = 0.173 \quad \lambda=0.86 \,\text{GeV}^{-2} \,.
\end{split}
\end{equation}

Regarding the neutron PDF, we use the same proton nonperturbative function and unpolarized PDF set but with the following substitution for the up and down quarks:
\begin{equation}
    u_n = d_p \,, \quad d_n = u_p \,, \quad  \bar u_n = \bar d_p \,, \quad \bar d_n = \bar u_p\, .
\end{equation}

In the following we show estimates for the transverse $\Lambda$ polarization integrated over its transverse momentum (or, more precisely, over $\bm{q}_T$), using Eq.~(\ref{eq:sidis_pol_ratio_int}). 

\begin{table}[h!]
\caption{Nucleon and electron beam energies and the corresponding c.m.~energy}
    \begin{tabular}{ccc}
    \hline 
    $E_{N}$ (GeV) & $E_{e}$ (GeV) & $\sqrt{s_{eN}}$ (GeV)\\
    \hline
    41 & 5 & 28.6\\
    100 & 10 & 63.2\\
    \hline
\end{tabular}
\label{tab:sep}
\end{table}

We will consider two different values of the c.m.~energy $\sqrt{s_{eN}}$,  Eq.~(\ref{eq:sidis_Q_sep}), reported in Tab.~\ref{tab:sep}, corresponding to various combinations of nucleon (electron) beam energies $E_N$ ($E_{e}$). We will keep fixed $y$ to $0.4$, and explore different values of $x_B$ and $z_{\Lambda}$. 

\begin{figure}[t]
{\includegraphics[trim =  0 50 0 120,clip,width=7.5cm]{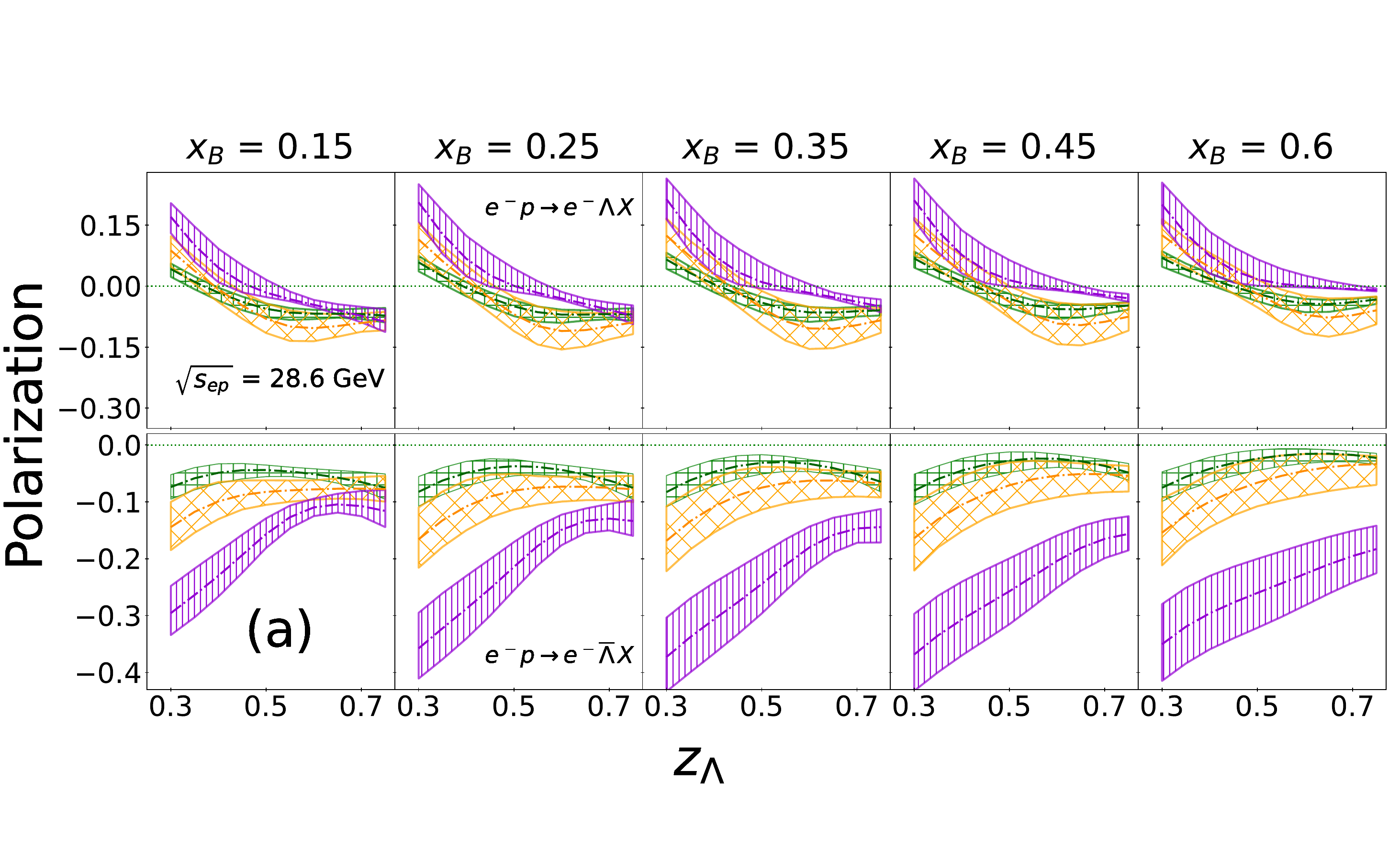}}
{\includegraphics[trim =  0 50 0 120,clip,width=7.5cm]{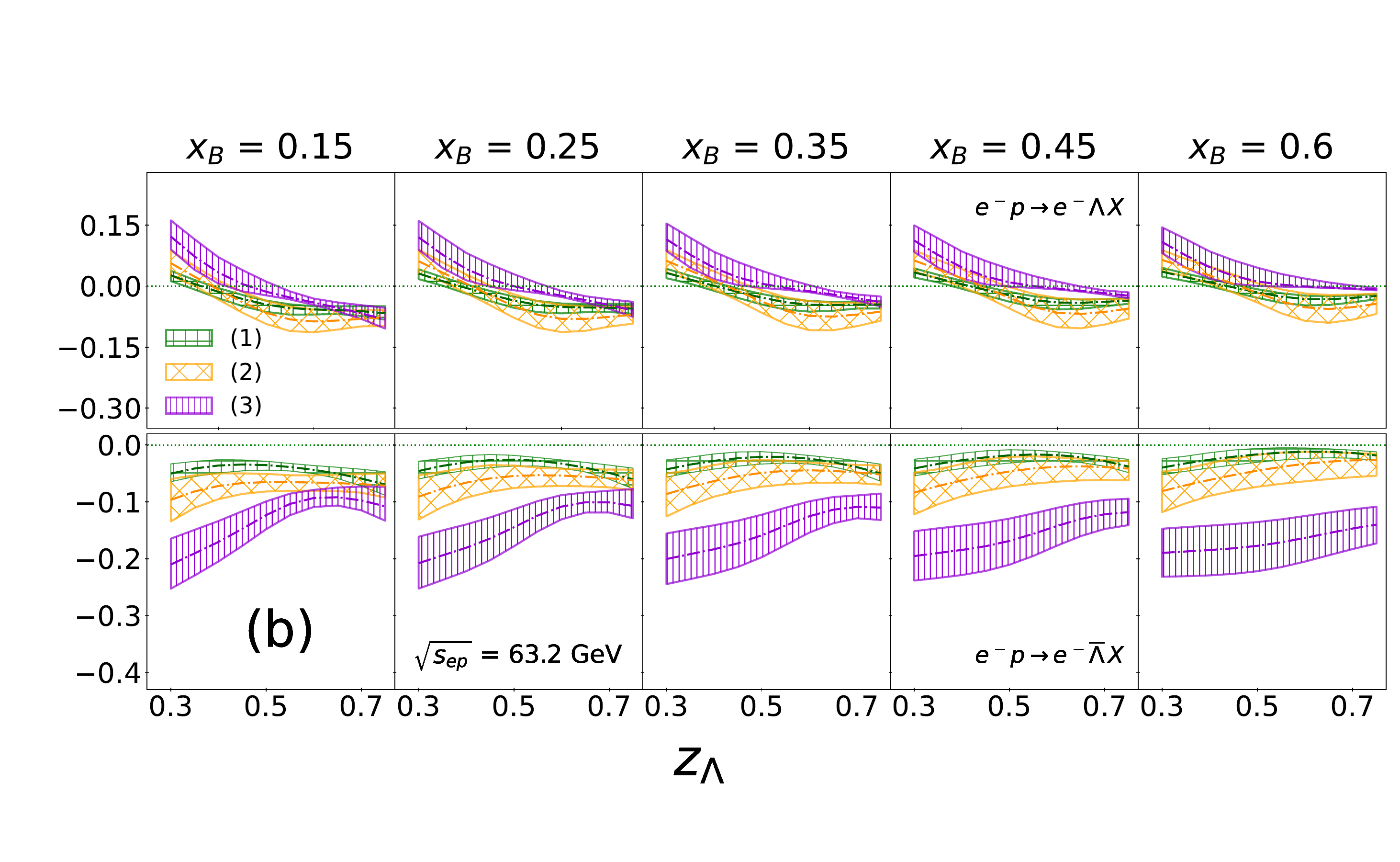}}
\newline
{\includegraphics[trim =  0 50 0 120,clip,width=7.5cm]{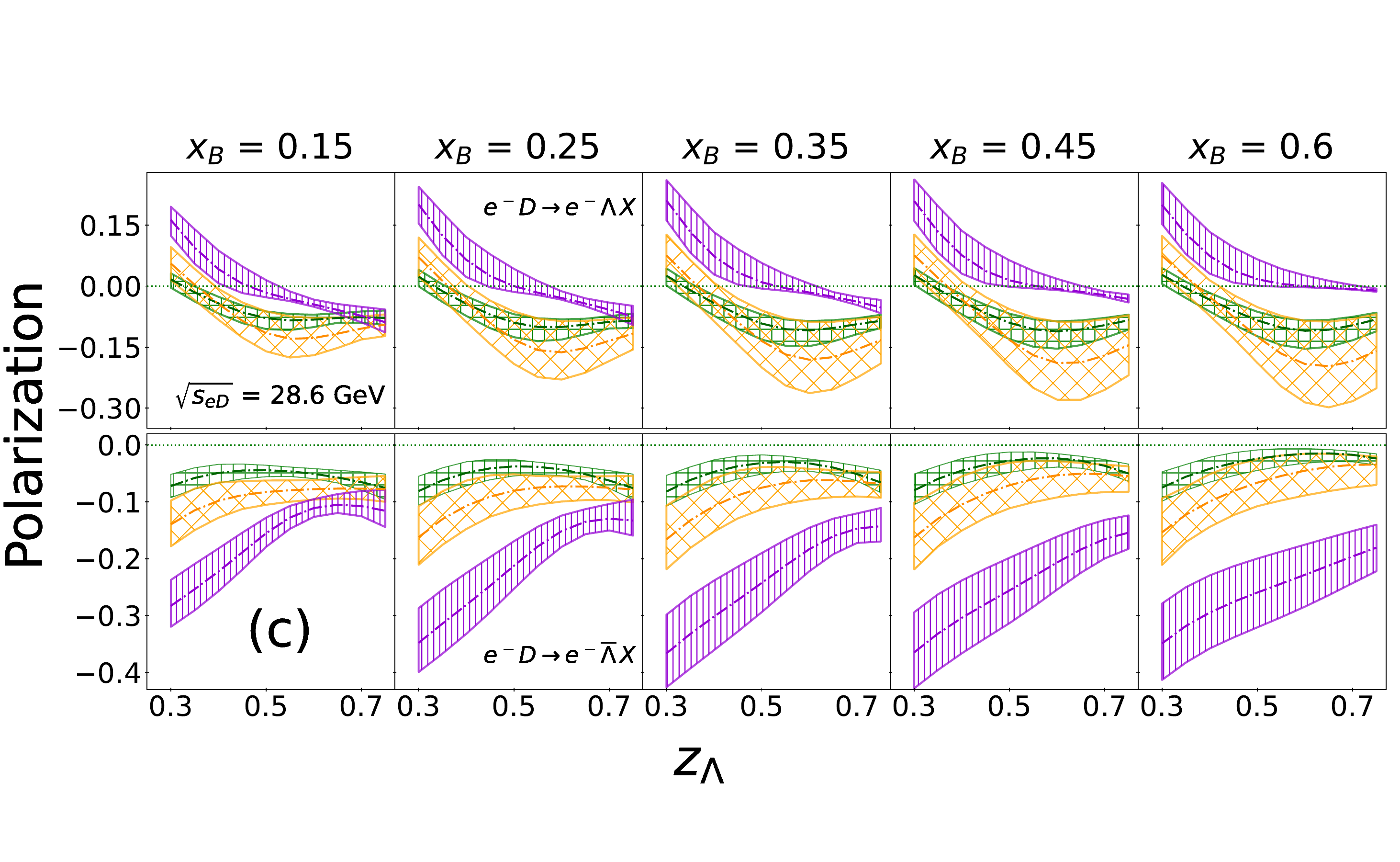}}
{\includegraphics[trim =  0 50 0 120,clip,width=7.5cm]{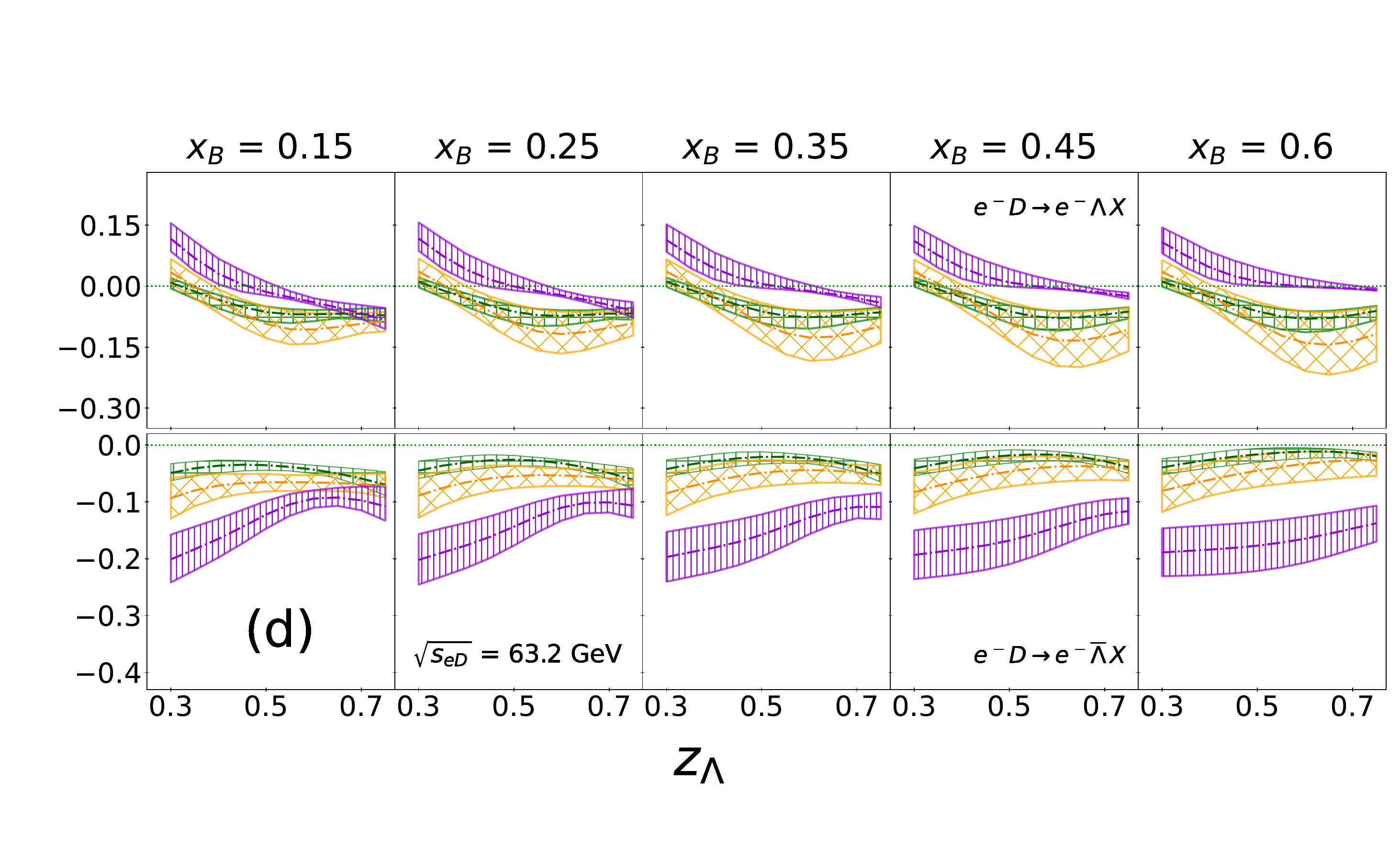}}
\caption{Estimates of the transverse $\Lambda/\bar{\Lambda}$ polarization in electron-proton (a, b) and electron-deuterium (c, d) scattering at $\sqrt{s_{eN}} = 28.6$~GeV (left panels) and 63.2~GeV (right panels), at $y=0.4$, as a function of $z_{\Lambda}$ for different $x_B$ bins, and the three scenarios: (1) green, (2) orange and (3) violet. The statistical uncertainty bands, at $2\sigma$ level, are also shown.}
\label{fig:sidis_pred_int} 
\end{figure}

In Fig.~\ref{fig:sidis_pred_int} we show the estimates for the $\bm{q}_T$-integrated transverse $\Lambda/\bar{\Lambda}$ polarization in electron-proton (deuterium) collisions, see Eqs.~(\ref{eq:sidis_pol_ratio_int}) and (\ref{eq:sidis_nuclei}), for different values of $\sqrt{s_{eN}}$, $x_B$ and $z_{\Lambda}$, in the three scenarios. 
Since we adopt a fixed ratio $q_{T_{\rm max}}/Q=0.27$, by exploring large values of $Q$, up to 30 GeV in our case, for certain $x_B$ values we enter the region of large $q_T$ (up to 8 GeV).

Firstly, we notice that scenarios 1 and 2 lead to $\Lambda/\bar{\Lambda}$ polarization with similar size and behavior, for the two values of $\sqrt{s_{eN}}$ both for proton and deuteron targets. 
%


We can see how the $\Lambda$ polarization tends to decrease, becoming negative, as $z_{\Lambda}$ increases, while the $\bar{\Lambda}$ polarization is always negative. In particular, for $\sqrt{s_{eN}} = 28.6$ GeV (Fig.~\ref{fig:sidis_pred_int}.a and \ref{fig:sidis_pred_int}.c), the polarization has the same pattern and size in each $x_B$ bin, while for greater $\sqrt{s_{eN}}$ values (Fig.~\ref{fig:sidis_pred_int}.b and \ref{fig:sidis_pred_int}.d), we have a general reduction in size of the polarization as $x_B$ grows.

For what concerns the third scenario, we can see that the polarization follows a pattern similar to that illustrated for the first and second scenarios, but with some differences. The $\Lambda$ polarization has a similar or slightly greater size than in the other two scenarios; the most significant difference can be found for the $\bar{\Lambda}$ polarization, which is much greater in size, reaching values of about $40\%$ for $x_B=0.6$ and $\sqrt{s_{eN}}=28.6\,\text{GeV}$. 
%

Finally we provide a comment on the strong similarities between the $\bar\Lambda$ polarizations in $ep$ and $eD$ collisions: the reasons can be traced back to the dominant contribution driven by the up and down distribution functions in both targets. 
For $\bar \Lambda$ production this enters 
directly convoluted with the polarizing FF for sea quarks in the numerator and the unpolarized sea FF in the denominator. 
In $\Lambda$ production this does not happen since the up and down parton distributions couple in a different way to the up and down pFFs when one considers a proton or a deuterium target.

At variance with the case of the double-hadron production in $e^+e^-$ collisions, the estimates for the transverse polarization within the second and third scenarios are clearly 
separated. Thus, future measurements of transversely polarized  $\Lambda/\bar{\Lambda}$ in SIDIS will potentially allow us to gain further insights and to distinguish between the two  scenarios.

It is worth noticing that the corresponding 
estimates for the transverse polarization as a function of the $\Lambda/\bar{\Lambda}$ transverse momentum, $P_{1T}$, are not able to discriminate among the different scenarios.

\subsection{Role of intrinsic charm contribution}

From the previous discussion it is clear that the charm contribution in the fragmentation process can be relevant for the study of the transverse $\Lambda$ polarization. 
Here, we explore how the employment of collinear PDFs, that take into account the presence of an intrinsic charm (IC) component in the proton, can play a role in this context. 
For this study, we consider again the CT14NNLO set and two recent PDF sets: the CT14NNLO IC set~\cite{Hou:2017khm}, by using the Brodsky-Hoyer-Peterson-Sakai (BHPS) model for the intrinsic charm component, and the NNPDF4.0 NNLO  set~\cite{NNPDF:2021njg}.

\begin{figure}[t]
{\includegraphics[trim =  0 0 0 0,clip,width=7.5cm]{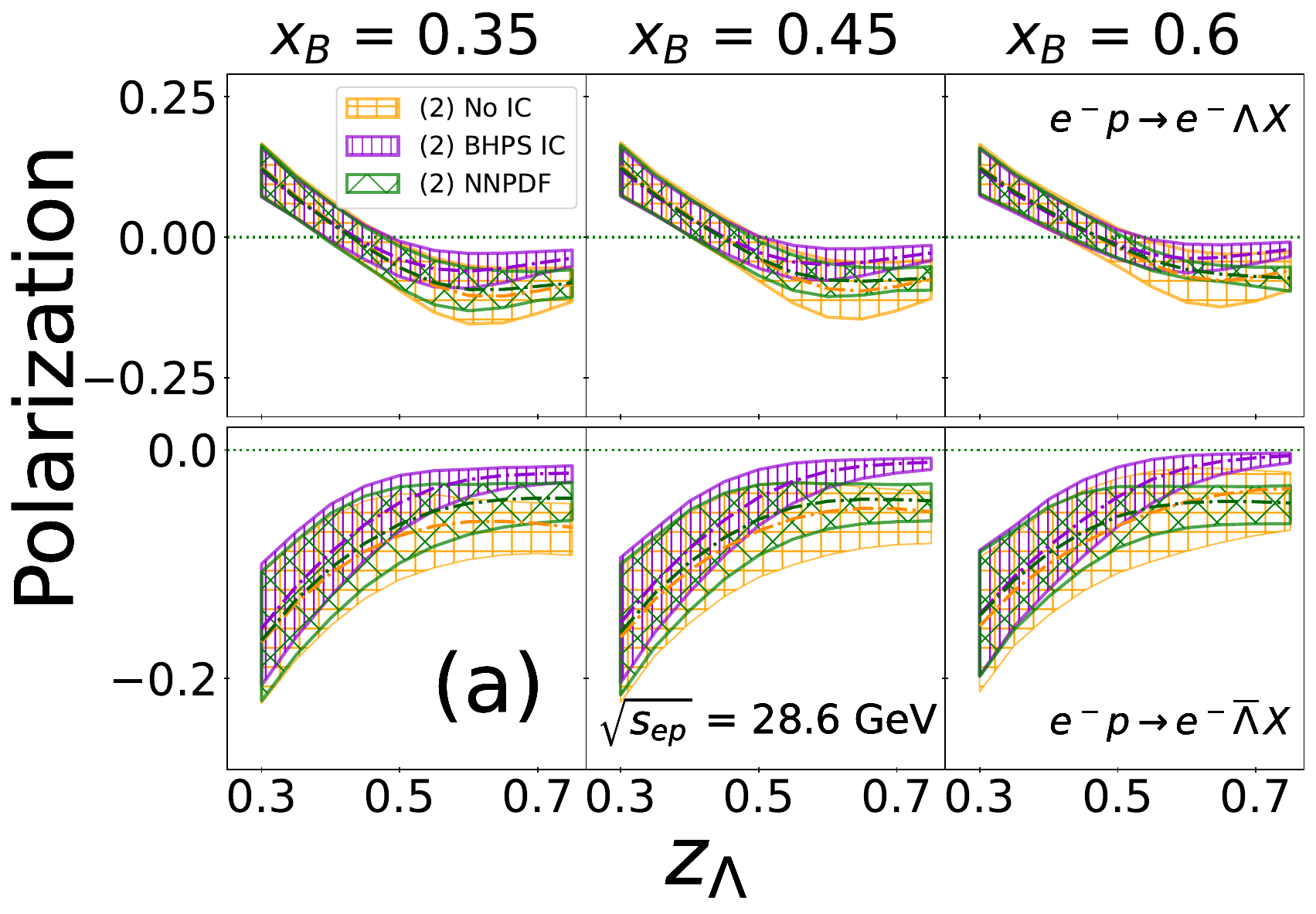}}
{\includegraphics[trim =  0 0 0 0,clip,width=7.1cm]{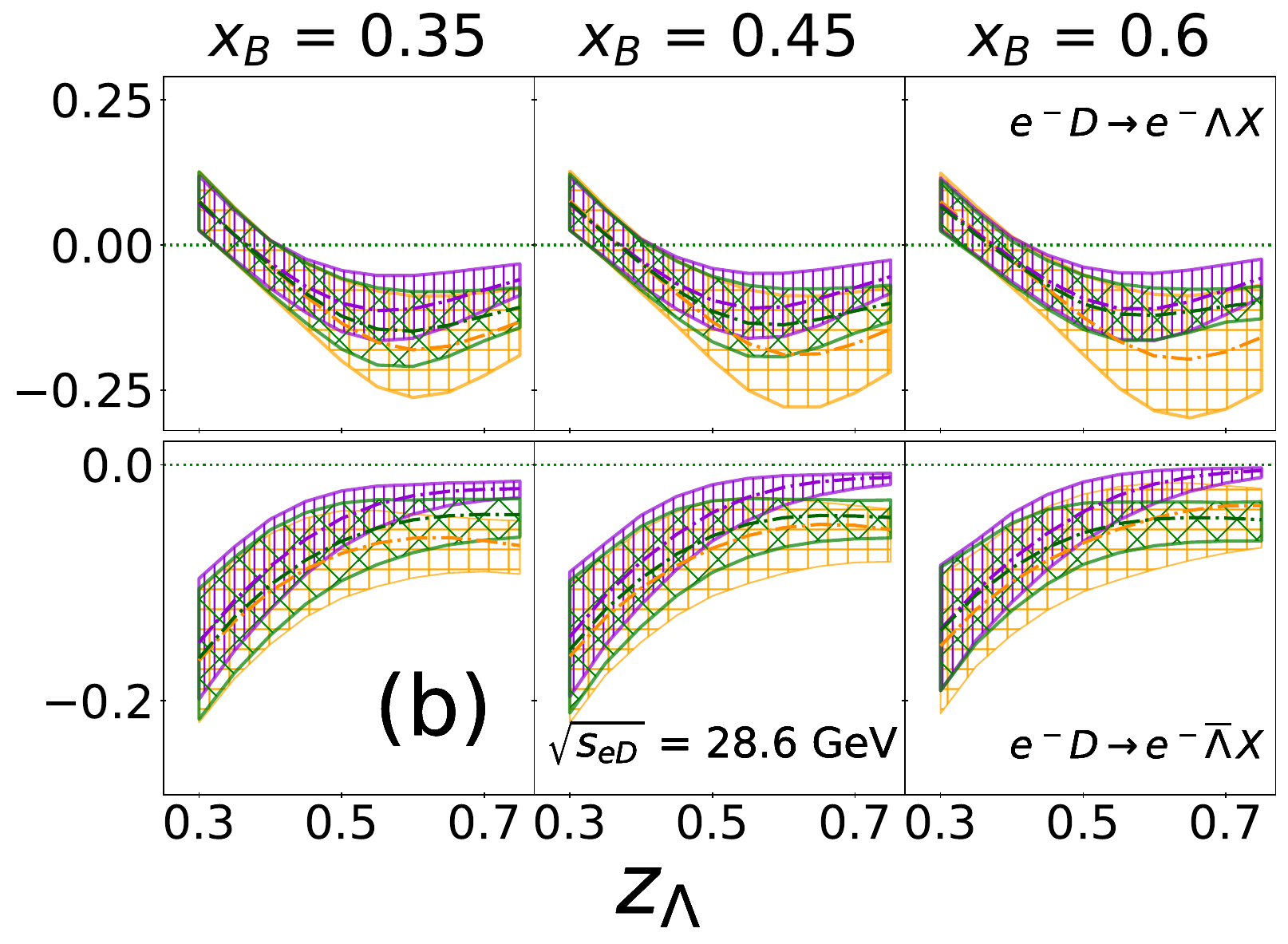}}
\caption{Estimates of the transverse $\Lambda/\bar{\Lambda}$ polarization in electron-proton (a) and electron-deuterium (b) scattering at $\sqrt{s_{eN}}= 28.6 \,\text{GeV}$, as a function of $z_{\Lambda}$ for different $x_B$ bins, obtained using the second scenario parameters for the first moment of the pFFs and different PDF sets: CT14NNLO no IC (orange bands), CT14NNLO IC with BHPS model for IC (violet bands) and NNPDF (green bands). The statistical uncertainty bands, at $2\sigma$ level, are also shown.}
\label{fig:sidis_IC_2_scen} 
\end{figure}

In Fig.~\ref{fig:sidis_IC_2_scen}, we present a comparison of the estimates of the transverse $\Lambda/\bar{\Lambda}$ polarization in electron-proton (deuterium) scattering at $\sqrt{s_{eN}} = 28.6 \,\text{GeV}$, obtained using the second scenario parameters for the polarizing FFs and the three PDF sets. We observe that the estimated polarization obtained using the BHPS model for IC (violet bands) and the NNPDF set (green bands) do not differ significantly from the predictions shown in the previous section (Fig.~\ref{fig:sidis_pred_int}) without the IC component (orange bands). This behavior is also present for smaller and greater values of the c.m.~energy. For completeness, we have also explored the role of the perturbative charm component (NNPDF set) without any significant differences.

\begin{figure}[t]
{\includegraphics[trim =  0 0 0 0,clip,width=7.5cm]{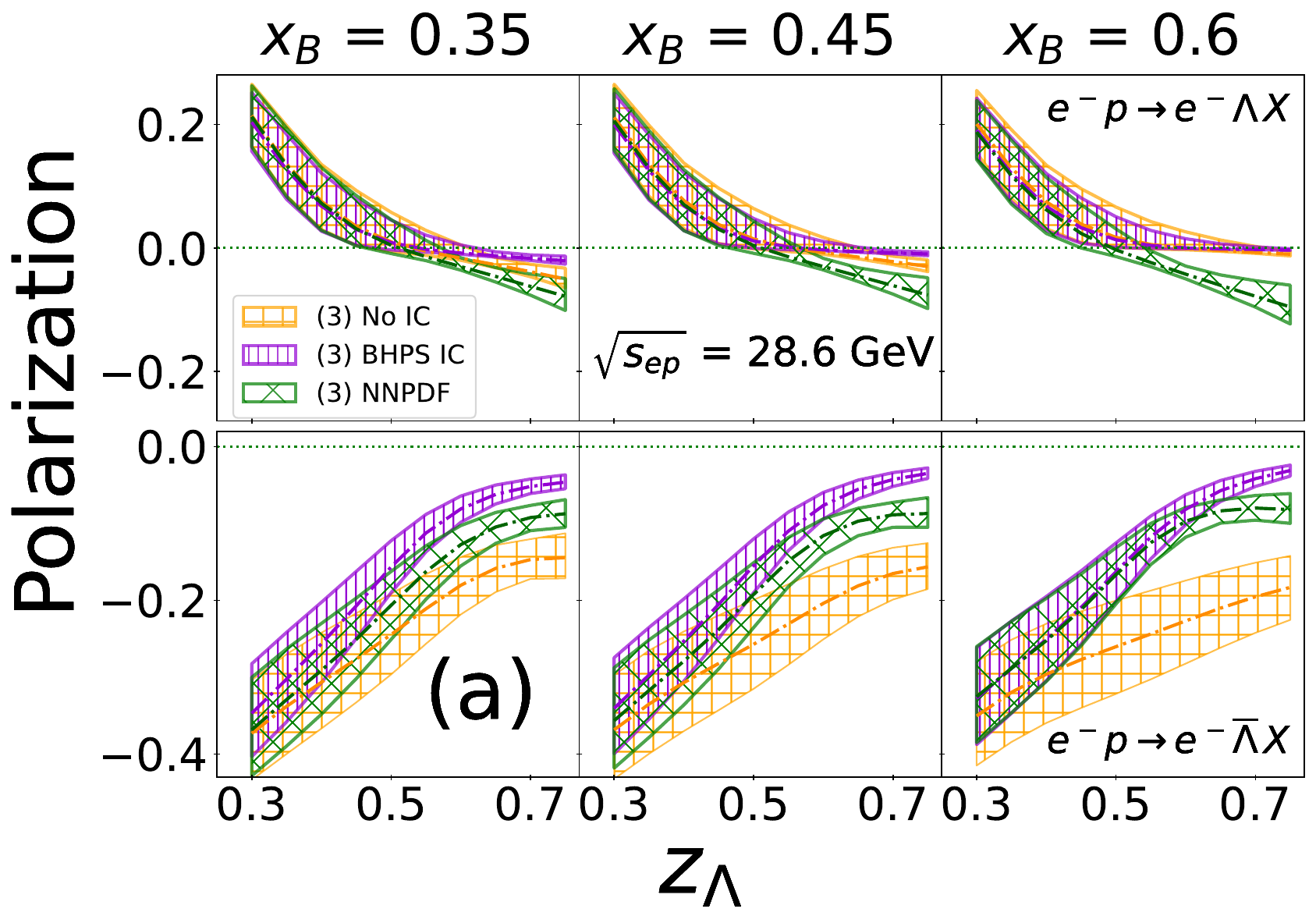}}
{\includegraphics[trim =  0 0 0 0,clip,width=7.cm]{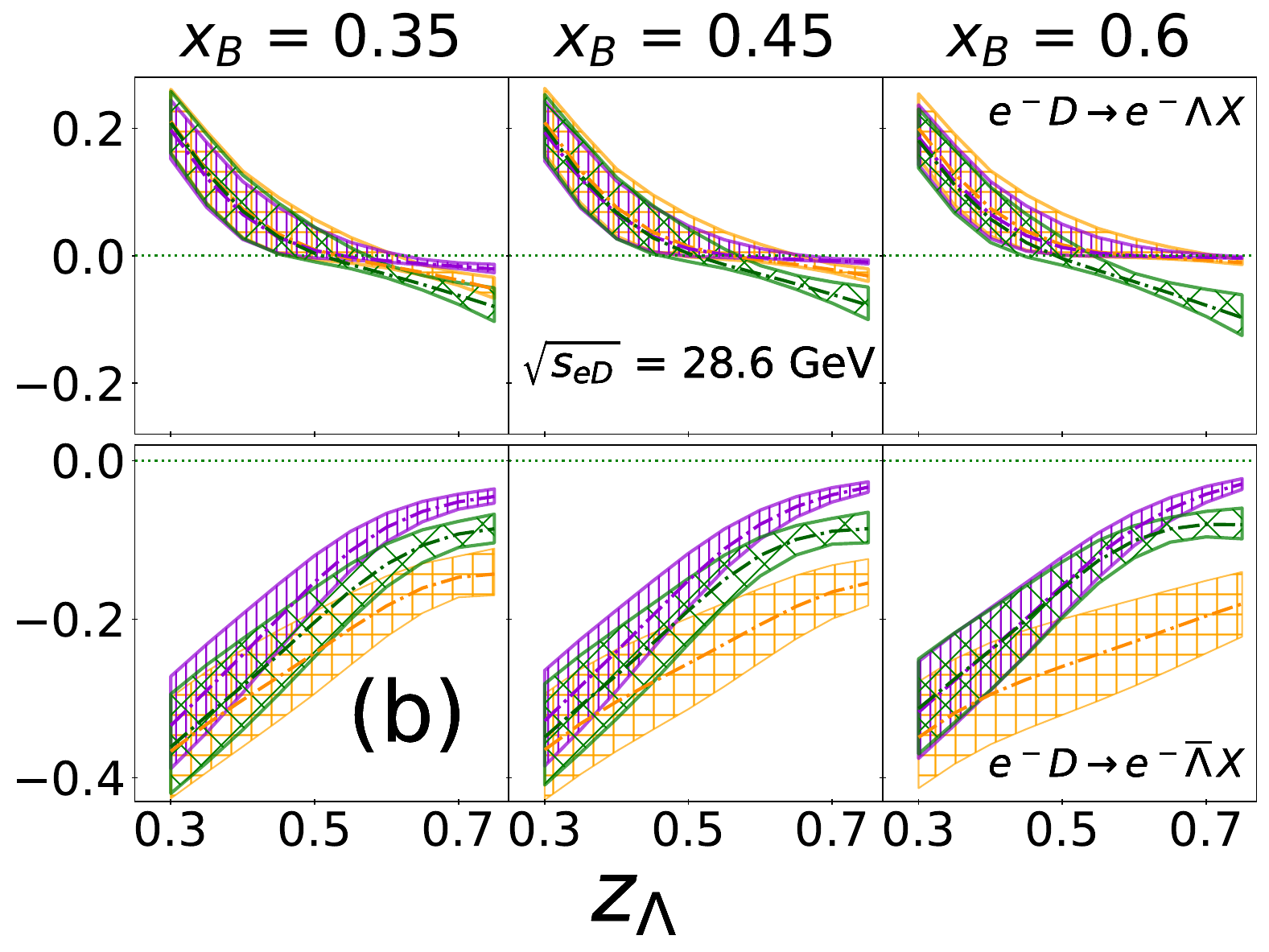}}
\caption{Estimates of the transverse $\Lambda/\bar{\Lambda}$ polarization in electron-proton (a) and electron-deuterium (b) scattering at $\sqrt{s_{eN}} = 28.6 \,\text{GeV}$, as a function of $z_{\Lambda}$ for different $x_B$ bins, obtained using the third scenario parameters for the first moment of the pFFs and different PDF sets: CT14NNLO no IC (orange bands), CT14NNLO IC with BHPS model for IC (violet bands) and NNPDF (green bands). The statistical uncertainty bands, at $2\sigma$ level, are also shown.}
\label{fig:sidis_IC_3_scen} 
\end{figure}

On the contrary, when adopting the third scenario parameters, the estimates can vary significantly as $x_B$ increases. As we can see in Fig.~\ref{fig:sidis_IC_3_scen}, the transverse $\bar{\Lambda}$ polarization estimates obtained with the BHPS model for IC (violet bands) and the NNPDF set (green bands) gradually move away from the predictions without the IC component, both in electron-proton and in electron-deuterium collisions, leading to a smaller polarization size. Notice that this conclusion is valid also when adopting the perturbative NNPDF charm component. 
Concerning the $\Lambda$ polarization, only the estimates with the NNPDF set differ from the other two predictions. In fact, both the bands without the IC component and the ones with the BHPS model decrease to zero, as $z_\Lambda$ increases, while the NNPDF predictions become negative and grow up to about $10 \%$ in size. It is worth noting that this is mainly due to the different PDF set adopted and not to the inclusion of the IC component.

\begin{figure}[t]
{\includegraphics[trim =  0 0 0 0,clip,width=11cm]{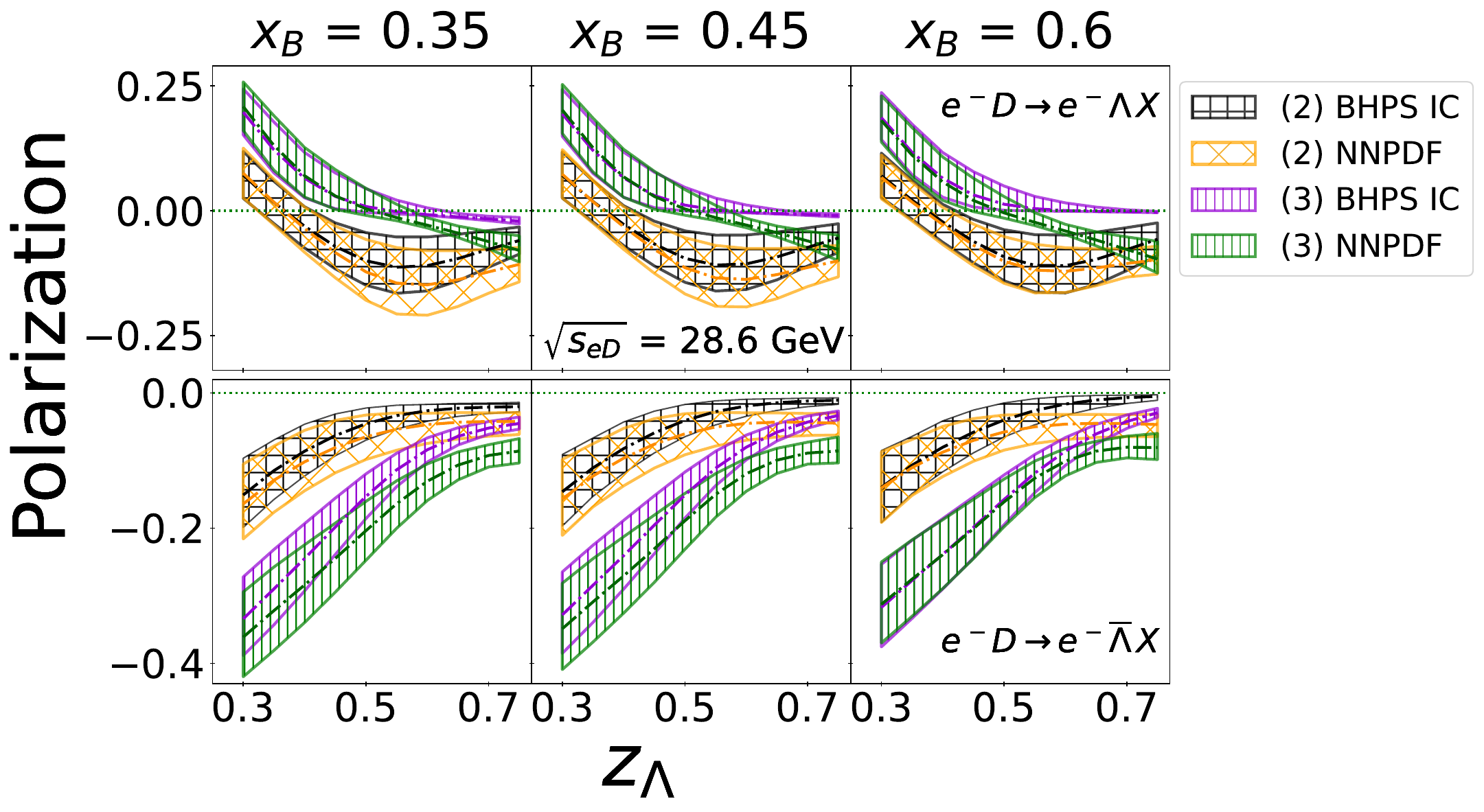}}
\caption{Estimates of the transverse $\Lambda/\bar{\Lambda}$ polarization and electron-deuterium scattering at $\sqrt{s_{eN}} = 28.6 \,\text{GeV}$, as a function of $z_{\Lambda}$ for different $x_B$ bins, obtained using the second and third scenario parameters for the first moment of the pFFs and different PDF sets: CT14NNLO IC with BHPS IC (black (Sc.~2) and violet (Sc.~3) bands) and NNPDF (orange (Sc.~2) and green (Sc.~3) bands). The statistical uncertainties, at $2\sigma$ level, are also shown.}
\label{fig:sidis_IC_23_scen} 
\end{figure}

In Fig.~\ref{fig:sidis_IC_23_scen}, we can see how, including the intrinsic charm contribution, the estimates for $\bar\Lambda$ in $eD$ collisions in the second and third scenarios are sufficiently well separated for both PDF sets.

\section{Conclusions}
\label{5_conclusions}

In this paper we have carried out a comprehensive reanalysis, within a TMD framework at NLL accuracy, of the transverse $\Lambda/\bar\Lambda$ polarization data from Belle Collaboration in the associated two-hadron production in $e^+e^-$ processes. In particular, we have focused on the role of isospin symmetry and of the charm contribution in the extraction of the polarizing fragmentation functions. While requiring $SU(2)$ simmetry alone (within a three-flavor scheme) leads to a very unsatisfactory fit, we have shown that all other scenarios considered allow for very similar and quite good descriptions of the available data. We can then conclude that Belle $e^+e^-$ data, or more generally $e^+e^-$ processes, alone are not able to discriminate among the different scenarios and, in particular, to shed light on the $SU(2)$ symmetry issue.

We have therefore explored this fundamental aspect by considering the same observable in SIDIS processes. By assuming the expected universality of the polarizing FFs we have given several predictions for the kinematical set-up reachable at the EIC, exploiting three different scenarios.
In such a case, by including the charm contribution in the unpolarized cross section, one can indeed distinguish between a scenario where isospin symmetry is respected or not. We have also considered a pFF for charm quarks in the numerator of the polarization without any improvement in the fit.
For completeness, we have discussed the role of the intrinsic charm in the proton for SIDIS processes and shown that the above conclusion does not change.

The spontaneous transverse $\Lambda$ polarization remains a challenging subject, but at the same time offers a unique opportunity to study the fragmentation mechanism and, more specifically, spin and transverse momentum correlations.
The present study, focused on processes where TMD factorization has been proven to hold, provides a further step to shed light on this very interesting phenomenon.

As we have shown, future EIC meaurements can play a significant role in this context: certainly in testing the phenomenological results obtained in $e^+e^-$ annihilation processes and more generally, in testing fundamental issues like the universality of the polarizing FFs, their scale dependence, their flavor decomposition a well as the role of $SU(2)$ symmetry.

\section*{Acknowledgments}
We thank Carlo Flore for his suggestions on the role of the intrinsic charm.  This project has received funding from the European Union’s Horizon 2020 research and innovation programme under grant agreement N.~824093 (STRONG-2020). U.D.~and M.Z.~also acknowledge
financial support by Fondazione di Sardegna under the projects ``Proton tomography at the LHC'', project number F72F20000220007 and ``Matter-antimatter asymmetry and polarisation in strange hadrons at LHCb`'', project number F73C22001150007 (University of Cagliari).
L.G.~acknowledges
support from the US Department of Energy under contract No.~DE-FG02-07ER41460. 
\appendix
\section{Perturbative Sudakov factor}
\label{Apx_A}

Here we give the analytic expression of the perturbative Sudakov factor presented in Eq.~(\ref{spert_def}):

\begin{equation}
    S_{\rm pert}(b_*;\bar{\mu}_{b})=-\widetilde{K}(b_*;\bar{\mu}_{b}) \ln\frac{Q^2}{\bar{\mu}_{b}^2} - \int^{Q}_{\bar{\mu}_{b}} \frac{d\mu'}{\mu'}\,\bigg[ 2\gamma_D(g(\mu');1) - \gamma_K(g(\mu')) \ln{\frac{Q^2}{\mu'^2}} \bigg] \,.
\end{equation}

As discussed in Section \ref{4_phenomenology}, since our goal is a phenomenological analysis at NLL accuracy, we take $\alpha_s$ at LO order:

\begin{equation}
        \alpha_s(\mu^2) = \frac{1}{\beta_0 \ln(\mu^2/\Lambda^2_{\rm QCD})}\,,
\end{equation}
and we expand the anomalous dimensions as follows:
\begin{equation}
    \gamma_K = \sum_n  \gamma^{[n]}_K \bigg(\frac{\alpha_s}{4\pi}\bigg)^n \, \quad \gamma_{D} = \sum_n  \gamma^{[n]}_{D} \bigg(\frac{\alpha_s}{4\pi}\bigg)^n \,,
\end{equation}
up to, respectively, the second and first order.
Given that the first order term of $\widetilde{K}(b_*;\bar{\mu}_{b})$ is zero~\cite{collins_2011,Aybat:2011zv}, the perturbative Sudakov factor can be written again as:

\begin{eqnarray}
      S_{\rm pert}(b_*;\bar{\mu}_{b})&=&\frac{\gamma^{[1]}_D}{4\pi \beta_0}\ln\bigg(\frac{\ln(Q/\Lambda_{\rm QCD})}{\ln(\bar{\mu}_{b}/\Lambda_{\rm QCD})}\bigg) 
    +\frac{\gamma^{[1]}_K}{4\pi \beta_0}\bigg[\ln(Q/\bar{\mu}_{b})- \ln(Q/\Lambda_{\rm QCD}) \ln\bigg(\frac{\ln(Q/\Lambda_{\rm QCD})}{\ln(\bar{\mu}_{b}/\Lambda_{\rm QCD})}\bigg) \bigg]\nonumber \\ 
    &+& \frac{\gamma^{[2]}_K}{2(4\pi \beta_0)^2} \bigg[- \frac{\ln(Q/\bar{\mu}_{b})}{\ln(\bar{\mu}_{b}/\Lambda_{\rm QCD})} + \ln\bigg(\frac{\ln(Q/\Lambda_{\rm QCD})}{\ln(\bar{\mu}_{b}/\Lambda_{\rm QCD})}\bigg)\bigg] \,,
\label{eq:spert_def_res}    
\end{eqnarray}
where~\cite{Collins:2017oxh}:
\begin{eqnarray}
    \beta_0&=&\frac{11 C_A - 4 T_F n_f}{12 \pi}\,, \qquad \gamma^{[1]}_{D}=6 C_F  \,,    \nonumber \\
    \gamma^{[1]}_K &=& 8 C_F \,, \qquad \gamma^{[2]}_K = C_A C_F \bigg(\frac{536}{9} -  \frac{8  \pi^2}{3}\bigg) - \frac{80}{9} C_F n_f \,,
\end{eqnarray}
with $C_F=4/3$, $C_A=3$, $T_F=1/2$, and $\Lambda_{\rm QCD}=0.2123 \,\text{GeV}$ for $n_f=3$ or $\Lambda_{\rm QCD}=0.1737 \,\text{GeV}$ for $n_f=4$.


\providecommand{\href}[2]{#2}\begingroup\raggedright\endgroup

\end{document}